\documentclass[aps,pre,amsmath,amssymb, twocolumn,showpacs,superscriptaddress, 10pt]{revtex4}
\usepackage{graphicx,bm,amssymb,amsmath,dcolumn,hyperref,epstopdf}
\usepackage{color,multirow}
\usepackage{mathrsfs}

\usepackage{float}
\usepackage{ulem}

\begin{document}
\definecolor{red}{rgb}{1,0,0}
\newcommand{\red}[1]{\textcolor{red}{#1}}

\newcommand*{\figuretitle}[1]{%
    {\raggedright
    \large
    \textbf{#1}
    \par\medskip}
}

\preprint{APS/123-QED}

\title{Insights into the Electron-Electron Interaction from Quantum Monte Carlo Calculations}


\author{Carl A. Kukkonen} 
\email{kukkonen@cox.net}
\noaffiliation{}

\author{Kun Chen}
\email{ kunchen@flatironinstitute.org }
\affiliation{Center for Computational Quantum Physics, Flatiron Institute, 162 Fifth Avenue, New York, NY 10010}

\date{\today}

\begin{abstract}
The effective electron-electron interaction in the electron gas depends on both the density and spin local field factors. Variational Diagrammatic Quantum Monte Carlo calculations of the spin local field factor are reported and used to quantitatively present the full spin-dependent, electron-electron interaction. Together with the charge local field factor from previous Diffusion Quantum Monte Carlo calculations, we obtain the complete form of the effective electron-electron interaction in the uniform three-dimensional electron gas. Very simple quadratic formulas are presented for the local field factors that quantitatively produce all of the response functions of the electron gas at metallic densities.

Exchange and correlation become increasingly important at low densities. At the compressibility divergence at rs = 5.25, both the direct (screened Coulomb) term and the charge-dependent exchange term in the electron-electron interaction at q=0 are separately divergent. However, due to large cancellations, their difference is finite, well behaved, and much smaller than either term separately. As a result, the spin contribution to the electron-electron interaction becomes an important factor. The static electron-electron interaction is repulsive as a function of density but is less repulsive for electrons with parallel spins.





The effect of allowing a deformable, rather than rigid, positive background is shown to be as quantitatively important as exchange and correlation. As a simple concrete example, the electron-electron interaction is calculated using the measured bulk modulus of the alkali metals with a linear phonon dispersion. The net electron-electron interaction in lithium is attractive for wave vectors $0-2k_F$, which suggests superconductivity, and is mostly repulsive for the other alkali metals.

\end{abstract}

\maketitle

\section{Introduction}
The electron-electron interaction is important for pairing and superconductivity, spin and magnetic phenomena, transport properties, and as an input for numerical calculations. Many interesting phenomena occur in exotic materials, some with reduced dimensionality and unusual topologies, and different theoretical approaches have been employed to explain experiment.

This paper looks back to the well-studied three-dimensional electron gas to examine the quantitative effects of exchange and correlation on the electron-electron interaction to see if any insight may be gained for problems of modern interest. 

The equation for the spin dependent effective electron-electron interaction in the three-dimensional electron gas is well-established in the mean field local approximation, and is given in terms of the local field factors that define all of the response functions of the electron gas \cite{ref1}.
Sum rules specify the local field factors at small and large wave vector $q$, and the difficult problem is to calculate the intermediate wave vector dependence. The Quantum Monte Carlo method is considered to produce accurate results.  The density local field factor was calculated many years ago by numerous methods including Diffusion Quantum Monte Carlo. Results for the spin local field factor using the Variational Diagrammatic Monte Carlo (VDMC) method were first published in 2019\cite{ref2} and additional results are presented here. The spin local field factor completes the specification of the effective electron-electron interaction and has motivated this paper. The local field factors are also known as exchange and correlation kernels in Time-Dependent Density Functional Theory.

The equation for the electron-electron interaction itself is indicative, but it is difficult to understand the relative importance of the terms until the actual local field factors are used to show quantitative results.

The numerically calculated values for the density and spin local field factors are discussed, and it is shown that they can approximated by very simple formulas that produce all of the response functions of the electron gas.

\section{Electron-electron Interaction}
Kukkonen and Overhauser\cite{ref1} (KO) demonstrated that the standard self-consistent perturbation theory based on Hartree-Fock theory and linear response theory could be used to calculate the effective many-body interaction $V_{ee}$ between two electrons in a simple metal, modeled by the electron gas, in terms of the density local field factor $G_+(q,\omega)$ and spin local field factor $G_-(q,\omega)$. 
\begin{eqnarray}
\label{KO}
    V_{e \vec{\sigma}_{1}, e \vec{\sigma}_{2}}	&=&
    \frac{4 \pi e^{2}}{q^{2}} \left(\frac{\left(\omega^{2}-\omega_{0}^{2}\right) /\left(\omega^{2}-\omega_{q}^{2}\right)}{\left(1-G_{+} Q\right)\left[1+\left(1-G_{+}\right) Q\right]}\right. \nonumber \\
    && \left. -\frac{G_{+}^{2} Q}{1-G_{+} Q}-\frac{G_{-}^{2} Q}{1-G_{-} Q} \vec{\sigma}_{1} \cdot \vec{\sigma}_{2}\right) 
\end{eqnarray}
This is the interaction to be used for calculating matrix elements between two electrons with momenta $k_1$ and $k_2$ and spins $\sigma_1$ and $\sigma_2$.  For parallel spins, the wave functions must be properly anti-symmetric.

The electron gas is characterized by the density parameter $r_s$, ($n=4\pi (r_s a_0)^3 \!/3$, $a_0$ is the Bohr radius). $Q = v\Pi^0$ where $v=4\pi e^2/q^2$ is the Coulomb potential and $\Pi^0(q,\omega)$ is the Lindhard function. For convenience, we will not usually explicitly present the wave vector and frequency dependence. We will also not explicitly use the word effective in discussing interactions. With the deformable background, the standard phonon frequencies and background (lattice) screening results were obtained and are represented by the frequency dependence of the first term in Eq. \eqref{KO}. The intuitive physics concepts behind the electron-electron interaction in Eq. \eqref{KO} are presented in Ref. \cite{ref1}. The frequencies $\omega_q$ and $\omega_0$ refer to the frequency response of the deformable background which is discussed in Ref. \cite{ref1} and Section VII of this paper.

The first calculations of the screened interaction in the electron gas were the Thomas Fermi interaction and its quantum mechanical extension by Lindhard\cite{ref3}. The Lindhard result is recovered by setting both $G$'s equal to zero which results in
\begin{eqnarray}
    V_{\text{Lindhard}} = \frac{4\pi e^2}{q^2 (1+Q)}
\end{eqnarray}
 In this approximation, the electron-electron, electron-test charge and test charge-test charge interactions are all the same. These latter interactions are discussed in Appendix B.

The KO electron-electron interaction has been verified by many-body calculations \cite{ref3-1}, and has been extended to multicarrier \cite{ref3-5}, spin polarized and two-dimensional systems\cite{ref3}. The potential for superconductivity without phonons using the KO interaction was examined by Takada\cite{ref4} and by Richardson and Ashcroft\citep{ref5}. Takada found superconductivity for a single carrier system and Richardson and Ashcroft did not. Richardson and Ashcroft stated that an important difference was that each group used different values of $G_+(q)$ and $G_-(q)$. Richardson and Ashcroft calculated their own $G$'s including frequency dependence. Takada used frequency independent $G$'s. The $G$'s considered here are the static local field factors.

Connecting with Feynman diagrams, the frequency independent factor in the first term can also be written as $\Lambda^2/\epsilon$ -- two vertex corrections divided by the dielectric function \cite{ref6}. The deformable background (lattice) screens the first term, the Coulomb interaction, but not the exchange and correlation and spin response terms which arise from summing ladder diagrams. 

For a rigid lattice, the frequency dependence in the first term of Eq. \eqref{KO} equals one, and the first term is repulsive. The second term is attractive. The spin dependent term is repulsive for opposite spins (singlet) and attractive for parallel spins (triplet). We initially consider the rigid background and discuss the deformable background in section VII below.

Quantitative evaluation of the electron-electron interaction is made in section  V using the values of $G_+(q)$ discussed in Appendix A and the new accurate values of $G_-(q)$ reported below.

\section{Spin Local Field Factor $\mathbf{G_-(q)}$}
A Quantum Monte Carlo method using renormalized Feynman diagrams was developed by Chen and Haule\citep{ref2}. The resulting Variational Diagrammatic Monte Carlo (VDMC) method is a generic many-body solver that was tested on the electron gas. The VDMC method was used to calculate the spin and density responses in the electron gas. It is well suited for finite temperatures. The description of the VDMC method, including grouping of Feynman diagrams, numerical approach and high precision results are given in Ref. \cite{ref2}.  

The data reported here are new VDMC calculations of the static spin local field factor with higher accuracy for densities $r_s = 1-4$ from $q = 0-2.34 \, k_F$, and additional results for $r_s=5$. The calculation temperature is $T = 0.025 \, T_F$ which is equivalent to $T = 0$.  The method provides the highest accuracy calculations of the $q=0$ susceptibility, and our calculations at finite $q$ also have high accuracy, but less than at $q=0$. Typical error bars are shown with the data. 
\begin{figure}[h!]
    \centering
    \includegraphics[width=1.0\columnwidth]{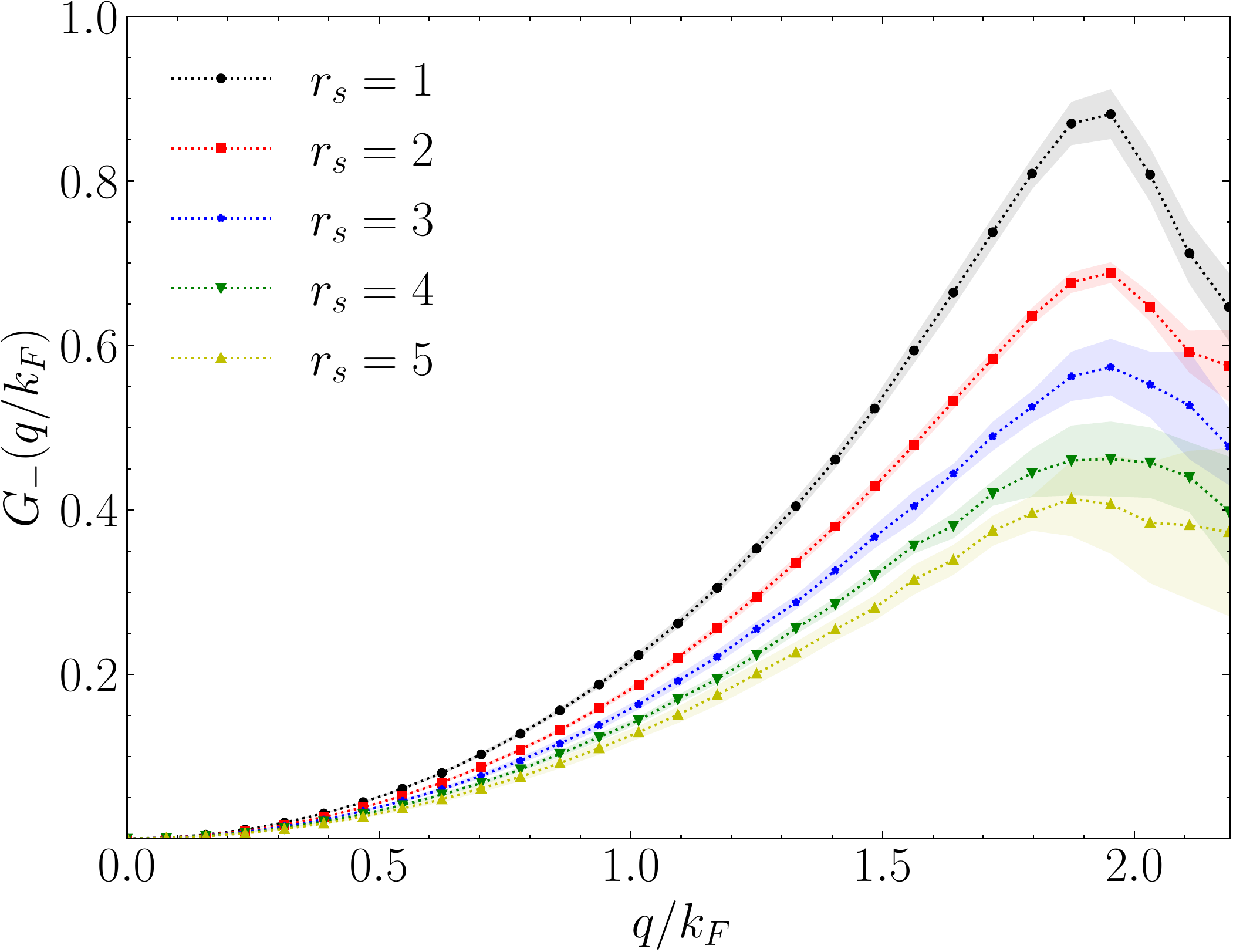}
    \caption{Spin local field factor $G_- (q)$ versus $q/k_F$ for $r_s= 1-5$ for the three-dimensional electron gas calculated with the Variational Diagrammatic Monte Carlo method. Error bars are shown by shading.}
    \label{Gminus}
\end{figure}

Figure \ref{Gminus} shows the wave vector dependence of $G_-(q)$ which demonstrates that it initially follows the quadratic behavior required by the susceptibility sum rule. The behavior changes dramatically near $q=2 \, k_F$. 

The susceptibility enhancement at $q=0$ was calculated more accurately than the values at finite $q$.  The results from Ref.\cite{ref2} and the new result at $r_s=5$ are reported in Table \ref{table1}.
\begin{table}[htbp]
	\begin{center}		
		\begin{tabular}{|c|c|c|c|c|c|}
			\hline
			$\mathbf{r_s}$	&	1	&	2	&	3	&	4	&	5						\\			\hline
			$\mathbf{\chi}/\mathbf{\chi_0} $	&	1.152(2)							&	1.296(6)	&	1.432(9)	&	1.576(9)							&	1.683(15)	\\	\hline
		\end{tabular}
		\caption{Susceptibility enhancement at $q=0$ for the three-dimensional electron gas calculated by Variational Diagrammatic Monte Carlo method. 					Uncertainty is indicated by the number in parentheses.}
		\label{table1}
	\end{center}
\end{table}

\begin{figure}[h!]
    \centering
    \includegraphics[width=1.0\columnwidth]{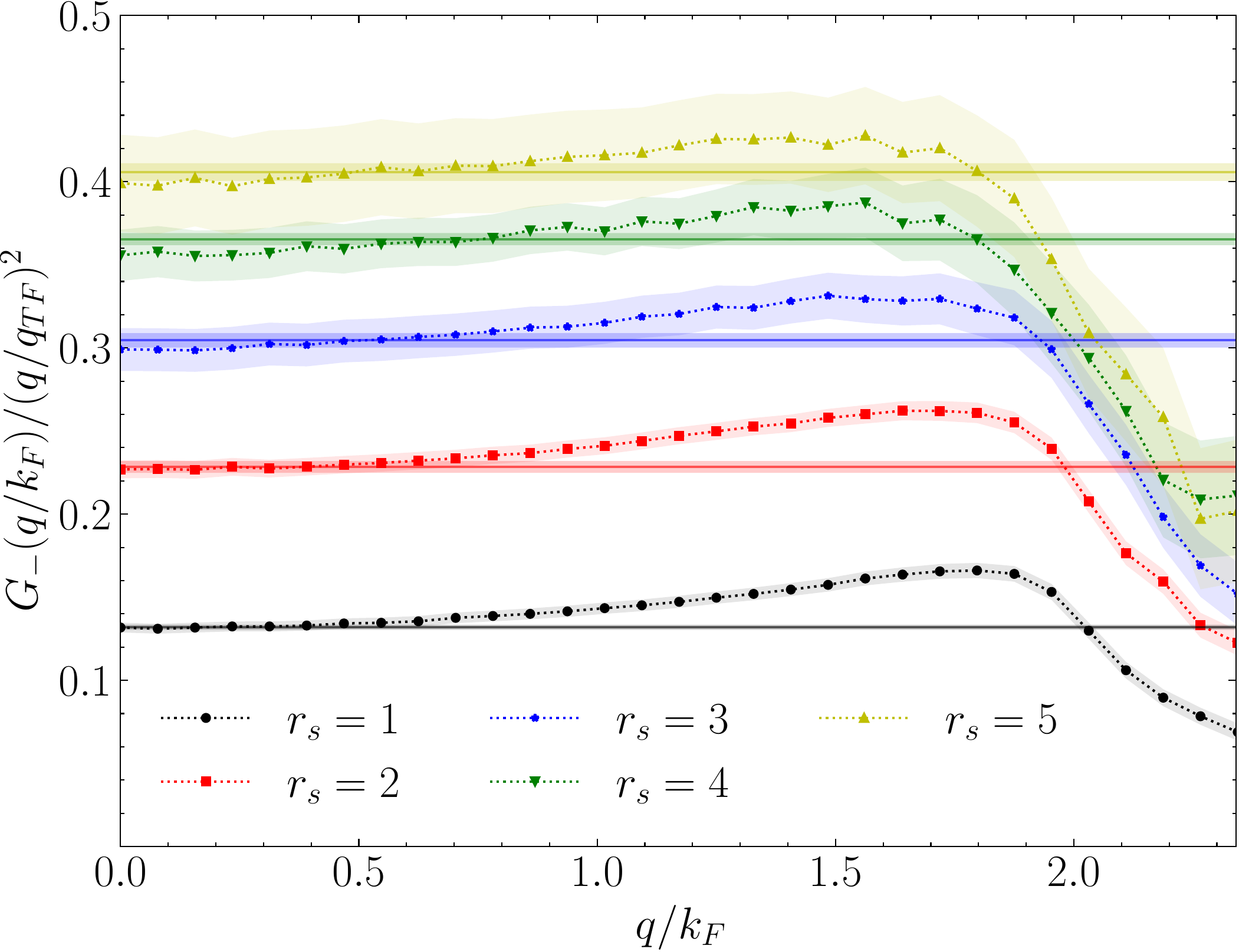}
    \caption{Spin local field factor divided by the wave vector divided by Thomas Fermi wave vector squared, $G_-(q)/(q/q_{TF})^2$ versus $q/k_F$ for 				$r_s = 1-5$. Error bars are shown by shading.}
    \label{ratio}
\end{figure}

In order to clearly see the small $q$ behavior and the effect of the susceptibility sum rule, the quantity $G_{-}(q)/(q/q_{TF})^2$ is plotted in Fig. \ref{ratio}.
The Thomas Fermi screening wave vector is defined by $q_{TF}^2=4k_F/\pi a_0$. The spin exchange and correlation kernel for Time Dependent Density Function Theory is $f_{xc}^{\text{spin}}=-4\pi G_-(q)/(q/q_{TF})^2$.

Figure \ref{ratio} shows that the spin local field factor $G_-(q)$ follows the quadratic well and does not fall below the quadratic behavior until near $2 \, k_F$. In fact, for high density $r_s=1$, $G_-(q)$ rises significantly above the quadratic before it falls below. The close adherence to the quadratic behavior, suggests that in the metallic region $r_s=2-5$, that a simple quadratic that satisfies the susceptibility sum rule is adequate to calculate the spin response function.

\begin{figure}[h!]	
	\centering
    \includegraphics[width=1.0\columnwidth]{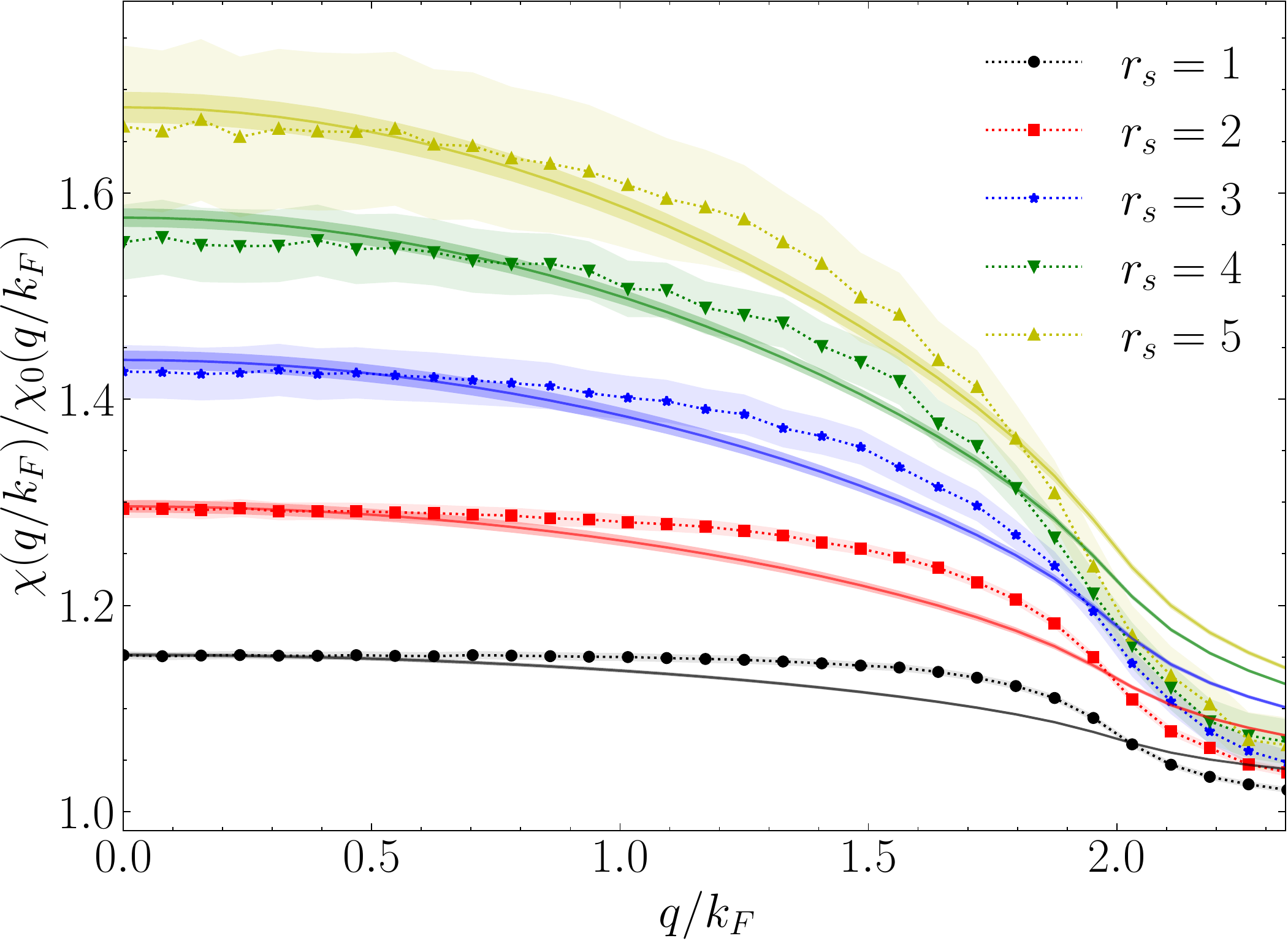}
    \caption{Susceptibility enhancement $\chi(q)/\chi_0$ plotted versus $q/k_F$ for $r_s=1-5$. The data points use the actual values of $G_-(q)$ 					calculated here and reported in Figure \ref{Gminus}. The solid lines are the simple quadratic function Eq. \eqref{eq5} set by the susceptibility sum rule at 				$q=0$. Error bars are shown by shading. Note that the $Y$ axis starts at 1.0.}
    \label{chi}
	\end{figure}

The wave vector dependent susceptibility enhancement is given by 
\begin{equation}
	\frac{\chi(q)}{\chi_0(q)} = \frac{1}{1-G_- Q}	\;	,
\end{equation}
and shown in Figure \ref{chi}.
	
The $q=0$ value of the susceptibility enhancement is entirely set by the susceptibility sum rule. The enhancement is modest because there is no divergence in the susceptibility near the metallic region. The simple quadratic, which is the horizontal line in Fig. \ref{ratio}, fits the data quite well and is adequate for the discussions in this paper and for comparison with experiment. If higher accuracy is needed, the actual data shown in Figure \ref{ratio} can be used.

The application of the VDMC method to the density local field factor $G_+(q)$ is briefly discussed in Appendix A.

\section{Simple Expressions for local field factors $\mathbf{G_+(q)}$ and $\mathbf{G_-(q)}$}
The recommended simple quadratic forms for $G_+(q)$ and $G_-(q)$ are:
\begin{equation}\label{eq4}
    G_+(q) = \left(1-\frac{\kappa_0}{\kappa}\right)\left( \frac{q}{q_{TF}}\right)^2
\end{equation}
\begin{equation}\label{eq5}
    G_-(q) = \left(1-\frac{\chi_0}{\chi}\right) \left(	\frac{q}{q_{TF}}	\right)^2 \; .
\end{equation}

These expressions are exact at small $q$ and accurately represent the QMC data up to almost $q=2 \, k_F$ for the metallic region $r_s=2-5$. $G_+(q)$ is discussed in Appendix A. Although these simple quadratic approximations are not accurate beyond $2 \, k_F$, they are suitable for the electron gas response functions which are cut off by the Lindhard function above $q=2 \, k_F$. For any application that requires values of $G$ at larger $q$, we recommend the interpolation formula discussed in Appendix A for $G_+$ or the actual data above for $G_-$.

The simple quadratic approximation to $G_-(q)$ given in Eq. \eqref{eq5} and used to calculate the susceptibility enhancement in Fig. \ref{chi} fits the VDMC data quite well. The fit is exact at $q=0$, falls below by $2\%$ at $q=1.5 \, k_F$ and slightly above at $2 \, k_F$. The average values are within $1\%$. The susceptibility is the product of the enhancement times the Lindhard function and the Bohr magneton. The falloff of the Lindhard function above $2 \, k_F$ makes this region unimportant for most applications. 

\begin{figure}[h!]	
    \centering
    \includegraphics[width=1.0\columnwidth]{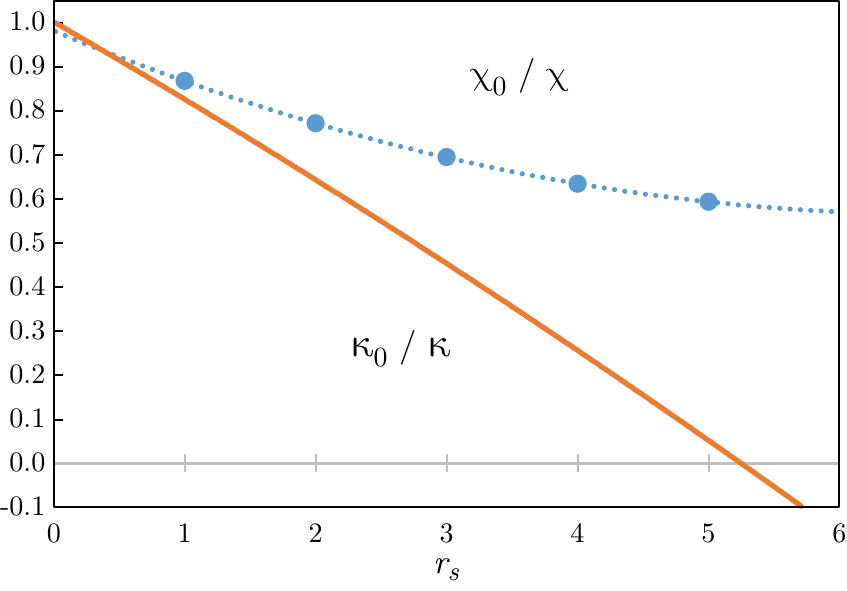}
    \caption{Compressibility ratio $\kappa_0/\kappa$ and susceptibility ratio $\chi_0/\chi$ for the three-dimensional electron gas with a rigid uniform 			positive background.}
    \label{kappa}
	\end{figure}

The recommended values of the compressibility are taken from Perdew and Wang\cite{ref7} and susceptibility ratios are given in Table \ref{table1}. Both are plotted in Figure \ref{kappa} and are accurately fitted by quadratic interpolation formulas. 

 The curves in Fig \ref{kappa} are fits to the data in the metallic region and fit the data to less than $0.5 \%$. Note that the compressibility and susceptibility ratios must equal 1 at $r_s=0$. Since we are only interested in the metallic region $r_s=1-5$, the fitting curves were not required to have an intercept of 1 at $q=0$, which results in a simpler and more accurate equations in the metallic region.
 
The compressibility and susceptibility ratios at $q=0$ are well fitted from $r_s=1-5$ by the following equations.
\begin{equation}
	\frac{\chi_0}{\chi} = 0.9821 - 0.1232 \, r_s + 0.0091 \, r_s^2
\end{equation}
\begin{equation}
	\frac{\kappa_0}{\kappa} = 1.0025 - 0.1721 \, r_s - 0.0036 \, r_s^2
\end{equation}
With these $G$'s and the compressibility and susceptibility ratios, all of the response functions for the three-dimensional electron gas with a rigid background can be quantitatively calculated. The same approach can be used for a spin polarized or two component electron gas.

Figure \ref{kappa} shows the well-known divergence and sign change of the compressibility at $r_s = 5.25$. This causes the vertex function and thus the dielectric function to diverge and become negative. The rigid uniform positive background prevents the overall model electron gas from becoming unstable. 

\section{Electron-Electron Interaction: Quantitative Results}
Using $G_+(q)$ and $G_-(q)$, we plot the electron-electron interaction. We plot each term in Eq. \eqref{KO} separately to show their relative importance. These are denoted $V_{ee}1$, $V_{ee}2$ and $V_{ee}3$. The first term $V_{ee}1$ is the coefficient of the frequency dependent factor (which is equal to $1$ for a rigid background). $V_{ee}1$ is intrinsically positive. The second term $V_{ee}2$ subtracts from the first term. $V_{ee}3$, the spin dependent term is positive (repulsive) for opposite spins (singlet) and negative (attractive) for parallel spins (triplet).

\begin{figure}[h!]
    \centering
    \includegraphics[width=1.0\columnwidth]{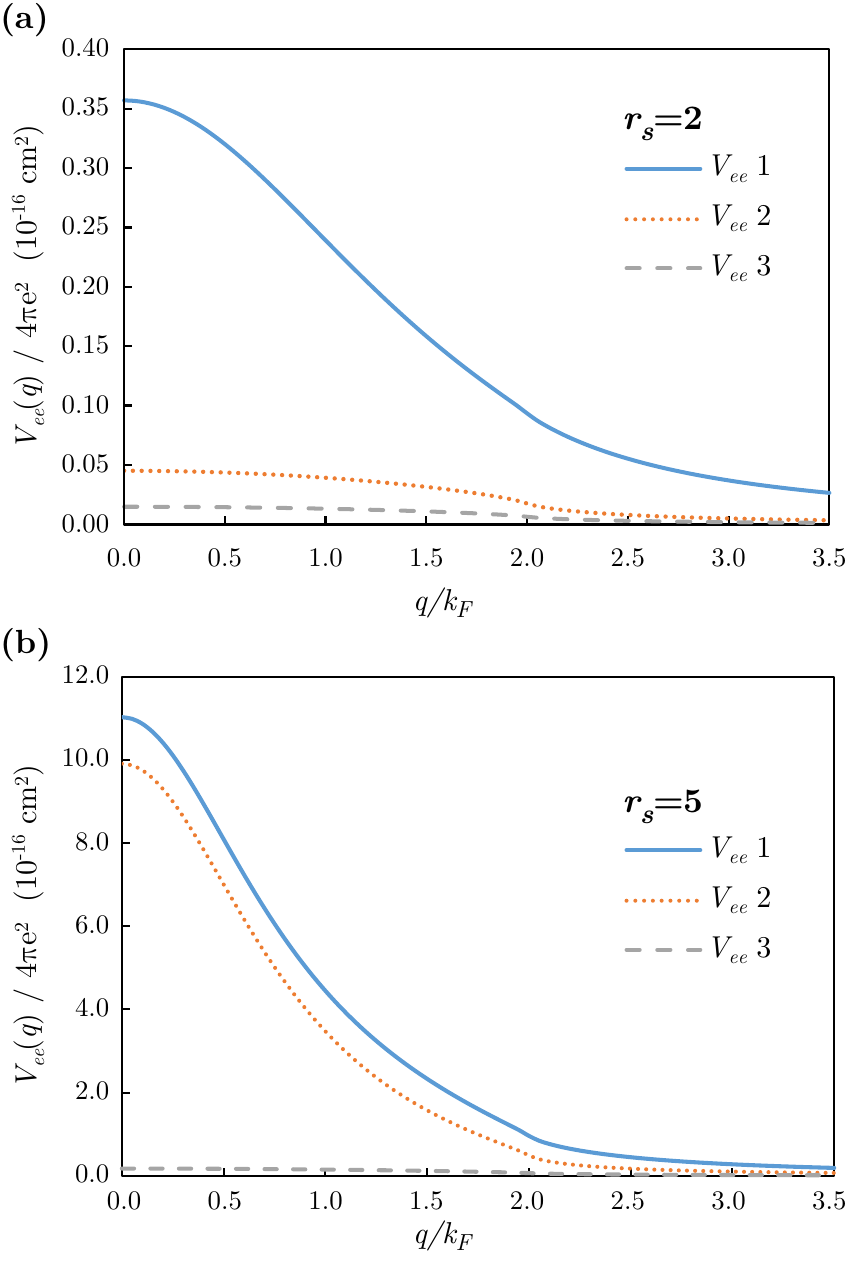}
    \caption{The three terms in the equation for $V_{ee}(q)$ for $r_s=2$ (a) and $r_s=5$ (b).  The first term $V_{ee}1$ is the screened Coulomb interaction. The second 		two terms are additional effects of exchange and correlation. The second term $V_{ee}2$ is subtracted from the first term and the third (spin dependent) term $V_{ee}3$ 		is subtracted for parallel spins and added for antiparallel (opposite) spins. The potential is measured in units of $4\pi e^2$.}
    \label{fig5}
	\end{figure}

These three terms are plotted together in Fig. \ref{fig5} for $r_s=2$ and $5$.

The magnitude of the first term $V_{ee}1$ at $q=0$ is $(\kappa/\kappa_0)/q_{TF}^{2}$ which is divergent at the compressibility divergence.  The results look ``normal" at $r_s=2$ where the second and third terms are small corrections to $V_{ee}1$ (the screened Coulomb interaction).  $V_{ee}1$ is screened by the deformable background (lattice) in the usual fashion. This seems consistent with the idea of perturbation theory where the corrections are small (except for the effect of the compressibility sum rule which is large even at $r_s=2$).  

At $r_s=5$, $V_{ee}1$ is very large. $V_{ee}1$ diverges as a function of $r_s$ at the compressibility divergence approximately as $1/(1-r_s/5.25)$. When calculating with Feynman diagrams, this term arises as the direct screened interaction with two vertex corrections 										$V_{ee}1=(\Lambda^2/ \epsilon) V_{ext}=\Lambda V_{et}$. This apparent divergence is one concern.

The second term $V_{ee}2$ is also large and apparently diverging, and is not screened by the lattice. $V_{ee}2$ is subtracted from $V_{ee}1$.  The fact that this term is large brings into question the use of perturbation theory and linear response.  However, upon closer examination, $V_{ee}2$ completely tracks $V_{ee}1$ and for a rigid background, they formally and exactly partially cancel each other to yield a finite value. This is due to a massive cancellation of Feynman diagrams using that approach.  For a rigid background, the difference $V_{ee}1- V_{ee}2 $ at $q=0$ is given as
\begin{equation}
	V_{ee}1(0) - V_{ee}2(0) = \frac{\kappa}{\kappa_0}\left( \frac{1-(\kappa/\kappa_0)^2}{q_{TF}^2} \right) \; = \; \frac{2-\kappa_0/\kappa}{q_{TF}^2}	\;	.
\end{equation}
For a rigid background, the net spin independent terms of the electron-electron interaction are completely well behaved and have no divergence at the compressibility divergence. This was also observed by KO\cite{ref1}.  The deformable lattice will be discussed in Section VII below.
	\begin{figure}[h!]	
    \centering
    \includegraphics[width=1.0\columnwidth]{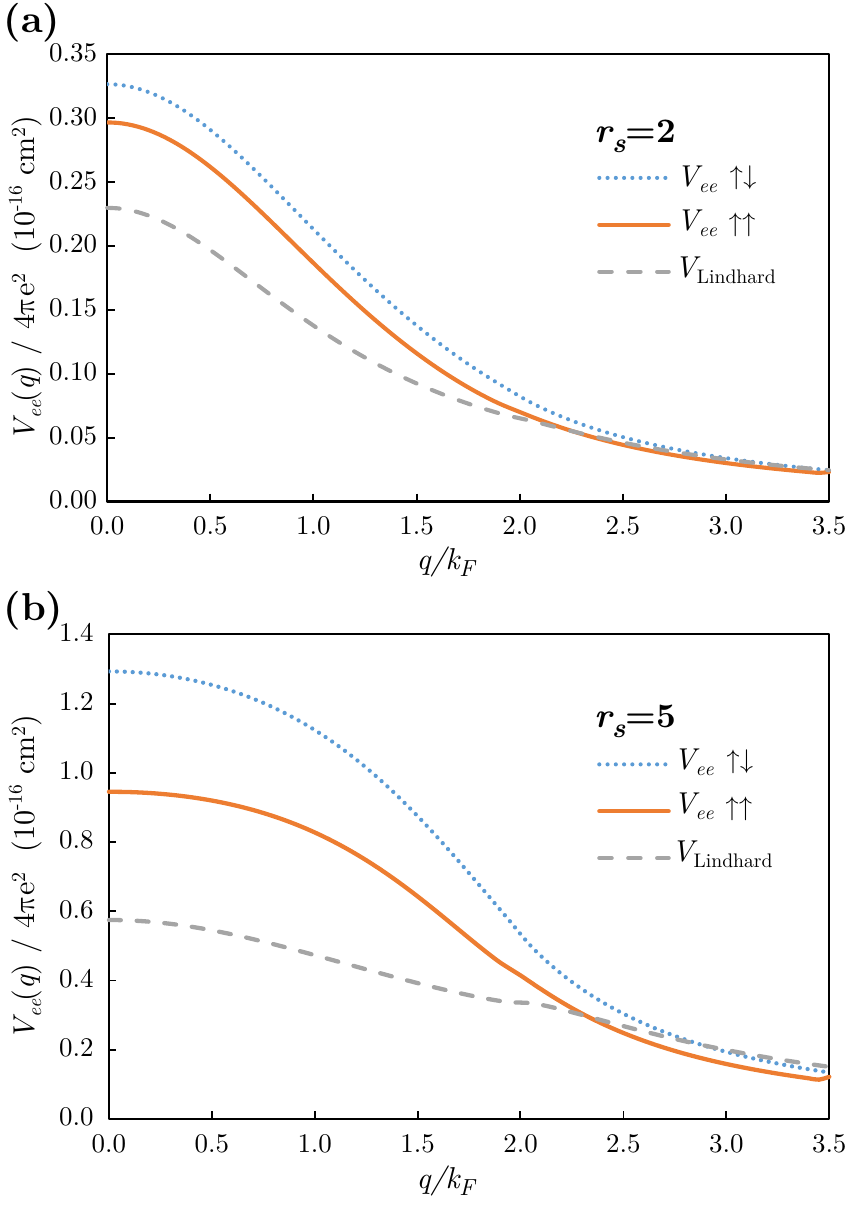}
    \caption{The net electron-electron interactions $V_{ee}(q)$ for opposite and parallel spins compared to the Lindhard potential at $r_s=2 $ (a) and $5$ (b). A 			rigid background is assumed for the electron gas.}
    \label{eeInteraction}
	\end{figure}

Figure \ref{eeInteraction} shows the net electron-electron interaction for electrons with parallel and opposite spins. This is the interaction to be used for calculating matrix elements and superconductivity. 
\begin{equation}
    V_{ee}^{\uparrow\downarrow} = V_{ee}1-V_{ee}2+V_{ee}3	
\end{equation}
is the interaction for opposite spins, and
\begin{equation}
    V_{ee}^{\uparrow\uparrow} = V_{ee}1-V_{ee}2-V_{ee}3
\end{equation}
is the interaction for parallel spins.

As mentioned above $V_{ee}1-V_{ee}2$ is a smooth function and lies midway between parallel and opposite spins. The overall electron-electron interaction is smooth and has no remaining evidence of the large effects in $V_{ee}1$ and $V_{ee}2$ individually. At $r_s=2$, $V_{ee}2$ and $V_{ee}3$ have effects at the $10 \%$ level. However at $r_s=5$, the first two terms nearly cancel, and $V_{ee}3$, the spin dependent term, is relatively important. The overall electron-electron interaction is considerably less repulsive for parallel spins. 

The values of the electron-electron interaction at $q=0$ are completely determined by the compressibility and susceptibility sum rules. At large $q$ (short distances), the electron-electron interaction follows the Lindhard function to the bare Coulomb interaction. The Lindhard interaction is shown for comparison.

This quantitative evaluation of the electron-electron interaction shows that with a rigid background, the static electron-electron interaction is well behaved and repulsive throughout the metallic region. The electron-electron interaction (and the other interactions in the electron gas discussed in Appendix B) are completely specified with simple equations provided in this paper. However, care must be taken with quantitative comparison with experiment. The effective mass, renormalization factor $z$, core polarization and a deformable lattice can have significant effects. Some of these were discussed by Kukkonen and Wilkins\cite{ref6}. The deformable lattice will be discussed in section VII.

\section{Effect of the Compressibility Sum Rule}
\begin{figure}[h!]	
    \centering
    \includegraphics[width=1.0\columnwidth]{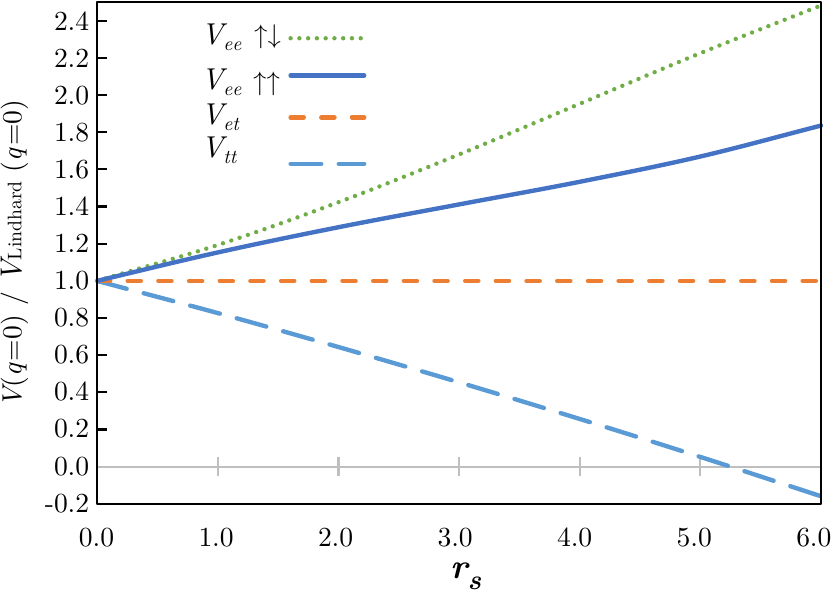}
    \caption{The ratio of the electron gas interactions at $q=0$ to the Lindhard (Thomas Fermi) interaction as a function of $r_s$ 			for a rigid background. The top curve is the electron-electron interaction for electrons with opposite spins. The next curve down is the electron-			electron interaction for parallel spins.  The electron-test charge interaction $V_{et}$ at $q=0$ is equal to the Thomas Fermi interaction at 		all $r_s$. The bottom curve is the test charge-test charge interaction $V_{tt}$ which always falls below Thomas Fermi interaction and becomes 				negative at the compressibility instability.}
    \label{ee|ee|vtt}
	\end{figure}

We have emphasized that the compressibility and susceptibility sum rules, which are derived from changes in the total electron gas energy, completely determine the various interactions at $q=0$. This is illustrated Fig. \ref{ee|ee|vtt} where we plot the ratio of the different interactions to the Lindhard interaction at $q=0$ versus $r_s$.

Figure \ref{ee|ee|vtt} shows that all of the interactions are the same and equal to the Lindhard and Thomas Fermi interactions at high density near $r_s=0$.  At lower density (larger $r_s$), the effects of exchange and correlation manifest themselves through the sum rules. Only the electron-electron interaction depends on spin and that is shown by the two curves for parallel and opposite spins.

With a rigid uniform positive background, the electron-electron and electron-test charge interactions show no unusual behavior as $r_s$ approaches the compressibility divergence at $r_s=5.25$.
However, the test charge-test charge interaction at $q=0$ becomes negative (attractive) above the instability. This is due to the fact that the dielectric function has become negative. This interaction is discussed in Appendix B.

The sum rules dictate the $q=0$ behavior. At large $q$, all the interactions fall off as $1/q^2$ which reflects the bare Coulomb interaction at short distances, and the wave vector dependence of the local field factors and the cut off of the Lindhard function at $2 \, k_F$  set the intermediate $q$ behavior which has been discussed above.

\section{Deformable Background}
KO considered a smooth but elastically deformable background which led to the frequency dependence in Eq.\eqref{KO}. The result was verified in Ref.\cite{ref3}. Introducing the deformable background naturally results in the phonon frequencies\cite{ref1}.
\begin{eqnarray}
\omega_{q}^{2} &=&\frac{N q^{2}}{M}\left(V_{i i}^{\text {bare }}-\frac{\left(V_{e i}^{\text {bare }}\right)^{2}}{v}+\frac{\left(V_{e i}^{\text {bare }}\right)^{2}}{v \epsilon}\right) \nonumber\\
&\equiv&\omega_{0}^{2}+\frac{N q^{2}\left(V_{e i}^{\text {bare }}\right)^{2}}{M v \epsilon}
\end{eqnarray}
Where $N$ is the density and $M$ is the mass of the background (ions), $V_{ii}^{\text{ bare}}$ is the bare ion-ion interaction, $V_{ei}^{\text {bare}}$ is the bare electron-ion interaction, and $\epsilon$ is the dielectric function. 

It is instructive to rewrite the frequency dependence in equation \eqref{KO} as
\begin{equation}
    \frac{\omega^2-\omega_{0}^{2}} {\omega^{2}-\omega_{q}^{2}}=1+\frac{\omega_{q}^{2}-\omega_{0}^{2}}{\omega^{2}-\omega_{q}^{2}}	\;	.
\end{equation}
This now multiplies the first term. The factor 1 allows the frequency independent part of the first term and second term to be formally combined which cancels the compressibility divergence in the static interaction, and Eq.\eqref{KO} can be rewritten as
\begin{eqnarray}\label{eq13}
    V_{e \vec{\sigma}_{1}, e \vec{\sigma}_{2}}&&=
    \frac{4 \pi e^{2}}{q^{2}} \left(\frac{\left(\omega^{2}_q-\omega_{0}^{2}\right) /\left(\omega^{2}-\omega_{q}^{2}\right)}{\left(1-G_{+} Q\right)\left[1+\left(1-G_{+}\right) Q\right]}\right.+ \nonumber \\
    && \left. \frac{1+(1-G_{+})G_{+} Q}{1+(1-G_{+}) Q}-\frac{G_{-}^{2} Q}{1-G_{-} Q} \vec{\sigma}_{1} \cdot \vec{\sigma}_{2}\right) 
\end{eqnarray}
The new first term $V_{ee-{\text{phonon}}}$ is divergent at 
$q=0$ at the compressibility divergence, but the second two terms have no divergence. 

$V_{ee-\text{phonon}}$ represents the additional screening of the Coulomb interaction by the background (lattice). At $\omega=0$, the numerator of the first term is negative which is the expected result for static screening by the positive background.
\begin{equation}\label{14}
    -\frac{\omega_q^2-\omega_0^2}{\omega_q^2} = \displaystyle{\frac{-N q^2 \left( V_{ei}^{\text{bare}}\right)^2		\Big/(M v e)}{(N q^2)/M	\displaystyle{\left(V_{ii}^{\text{bare}}			- \frac{\left( V_{e i}^{\text {bare }} \right) }{v} + \frac{\left( V_{e i}^{\text {bare }}\right)^2}{\epsilon \, v } \right)}}}
\end{equation}
The screening depends on the stiffness of the background represented by $V_{ii}^{\text{bare}}$ and the properties of the electron gas. The rigid background is obtained when $V_{ii}^{\text{bare}}$ (and thus $\omega_q^2$) goes to infinity and this term goes zero. 

Another interesting limit occurs if all of the interactions including $V_{ii}^{\text bare}$ are Coulomb interactions. In this case
\begin{equation}
    -\frac{\omega_q^2-\omega_0^2}{\omega_q^2}=-1
\end{equation}
and $V_{ee-\text{phonon}}$ is large and negative (attractive) At $q=0$, the first term has the value set by the compressibility sum rule as
\begin{eqnarray}
	V_{ee-\text{phonon}}(q=0) &=& -\frac{\kappa}{\kappa_0} V_{et}(q=0) \nonumber	\\
		&=& - \frac{\kappa}{\kappa_0} V_{TF}(q=0)
\end{eqnarray}
With all Coulomb interactions, this negative first term is larger than the other two terms combined and the overall electron-electron interaction at $q=0$ is attractive. Note this term diverges at the compressibility divergence.

An instructive intermediate case is to consider that $V_{ei}^{\text{bare}}$ in the numerator of Eq.\eqref{14} is equal to the Coulomb interaction $v=4\pi e^2/q^2$, and to take $\omega_q$ from experiment. The background (lattice) screening term can be rewritten as 
\begin{eqnarray}
    V_{ee-\text{phonon}}(q) &=& - \frac{1}{\omega_q^2} \left( \frac{v}{(\epsilon(1-G_{+}Q))^2} \right) \frac{Nq^2}{M} \nonumber	\\
    &=& - \frac{1}{\omega_q^2} V_{et}^2(q) \frac{N q^2}{M}
\end{eqnarray}
All of the effects of the deformable background are in $\omega_q^2$, the phonon frequencies which depend on the electron gas parameters as well as the bare ion-ion interaction. The electron test charge interaction appears because the lattice appears as a test charge to the electron gas. The electron test charge interaction is equal to the Thomas Fermi and Lindhard interactions at $q=0$, but differs at larger q as shown in Appendix B.

As the simplest example, we model the phonons by the relationship between $\omega_q$ and the bulk modulus B which is the inverse of the compressibility $\kappa$. 
\begin{equation}\label{eq18}
	\omega_q^2=\frac{Bq^2}{N M}
\end{equation}
where $NM$ is the mass density of the background, and $\sqrt{B/NM}$ is the speed of sound. Using this relationship and the bulk modulus of the non-interacting electron gas $B_0=1/(n^2 V_{TF}(q=0))$, and the fact that the ion density and the electron density are the same for the monovalent alkali metals, one obtains the equation for the background screening contribution to the electron-electron interaction
\begin{equation}\label{Vee|phonon}
    V_{ee-\text{phonon}}(q) = -\frac{V_{et}(q))^2}{V_{et}(0)} \frac{B_0}{B_{\text{experiment}}}
\end{equation}

The measured bulk moduli, free electron values\cite{ref8}, and their ratios for the alkali metals are given in Table \ref{table2}.
\begin{table}[htbp]
\begin{center}
    \begin{tabular}{|c|c|c|c|c|}
\hline & $r_s$ & $B_0$ & $B_{\text{experiment}}$ & $B_0/B_{\text{experiment}}$ \\
\hline Li & 3.25 & 23.9 & 11 & 2.173 \\
\hline Na & 3.93 & 9.23 & 6.3 & 1.465 \\
\hline K & 4.86 & 3.19 & 3.1 & 1.029 \\
\hline Rb & 5.2 & 2.28 & 2.5 & 0.912 \\
\hline Cs & 5.62 & 1.54 & 1.6 & 0.963 \\
\hline
\end{tabular}
\caption{Free electron and experimentally measured bulk moduli for the alkali metals.}
\label{table2}
\end{center}
\end{table}
	\begin{figure}[h!]		
    \centering
    \includegraphics[width=1.0\columnwidth]{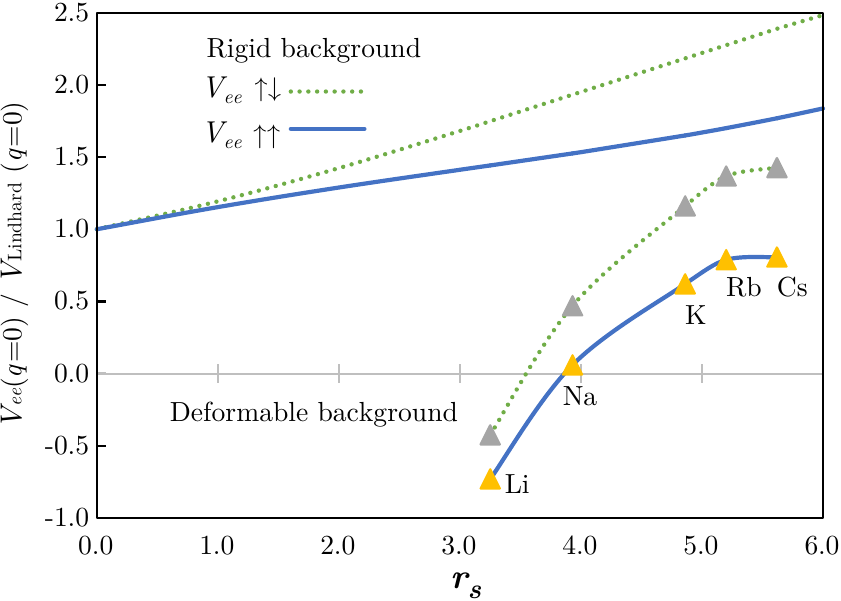}
    \caption{Static $(\omega=0)$ electron-electron interaction at $q=0$ for the electron gas at the density of alkali metals (in units of the Lindhard 				(Thomas Fermi) interaction). The top two curves are the electron-electron interaction for a rigid background with opposite spins, and parallel spins. 			The data at the lower right are the electron-electron interaction including the deformable background screening for the alkali metals using the 			measured bulk modulus for opposite and parallel spins. }
    \label{staticomega|ee|}
	\end{figure}
	
In Figure 8, the repulsive electron-electron interactions at $q = 0$ for opposite and parallel spins in a rigid background is plotted versus $r_s$ as in Figure \ref{ee|ee|vtt}. Also plotted is the net electron-electron interaction including screening by the deformable background using the experimentally measured bulk modulus for the alkali metals.	
	
The electron-electron interactions for a rigid background are simply the second two terms of Eq. \eqref{eq13} evaluated at $q = 0$. These are completely specified by the compressibility and susceptibility sum rules, and the values are given in Fig. \ref{kappa}.
\begin{equation}
	V_{ee}^{\uparrow \downarrow} (0) = \left[\left(	2-\frac{\kappa_0}{\kappa}	\right) + \frac{\chi}{\chi_0} \left(	1- \frac{\chi_0}{\chi}	\right)^2		\right] V_{\text{Lindhard}}(0)
\end{equation}
\begin{equation}
	V_{ee}^{\uparrow \uparrow} (0) = \left[\left(	2-						\frac{\kappa_0}{\kappa}	\right) - \frac{\chi}{\chi_0} \left(	1-	\frac{\chi_0}{\chi}	\right)^2		\right] V_{\text{Lindhard}}(0)
\end{equation}
The attractive background screening contribution to the electron-electron interaction is given by Eq. \eqref{Vee|phonon} evaluated at $q = 0$. The electron-test charge interaction $V_{et}(q)$ is given in Eq. \eqref{B3} in Appendix B, and is equal to the Lindhard and Thomas Fermi potentials at $q = 0$.
\begin{equation}
	V_{et - \text{phonon}}(0) = \left[	\frac{B_0}{B_{\text{experiment}}}	\right] V_{\text{Lindhard}}(0)
\end{equation}
This term is spin independent and is subtracted from both the opposite and parallel spin electron-electron interactions for a rigid background and net electron-electron interaction is plotted at the $r_s$ values of the alkali metals.
\begin{figure}	
    \centering
    \includegraphics[width=1.0\columnwidth]{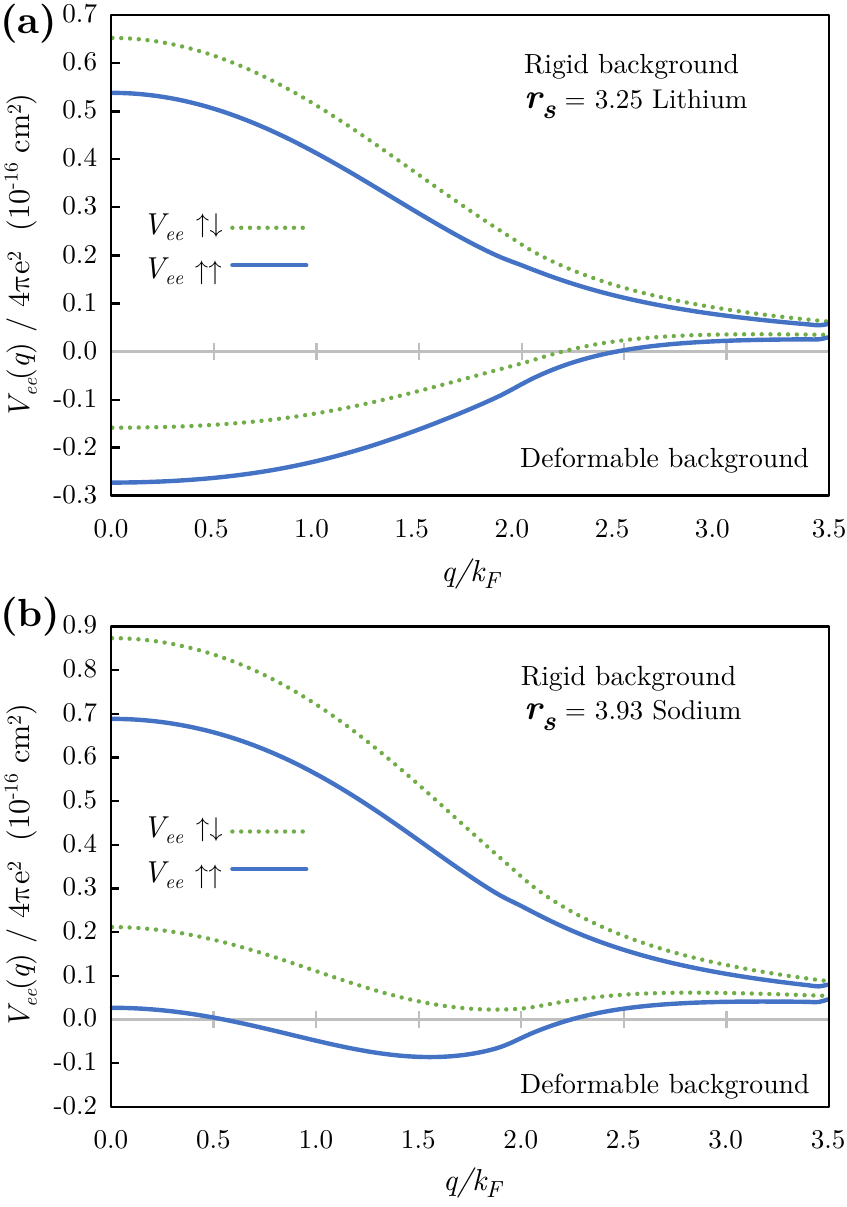}
    \caption{Electron-electron interaction at densities of lithium (a) and sodium (b) comparing the repulsive interaction calculated for a rigid background to the 			net interaction when the deformable background is included using the measured bulk modulus. The top two curves are for the rigid lattice for 				opposite and parallel spins.  The bottom two curves subtract the deformable background contribution from the top two curves. They represent the net 		electron-electron interaction. }
    \label{ee|Li,Na}
	\end{figure}
	
Note that all of the electron-electron interactions at $q = 0$ are proportional to $V_{\text{Lindhard}} (q=0) = 1/q_{TF}^2 = r_s \, a_0 / (2.95)^2$ in units of $4\pi e^2$.  The electron-electron interactions divided by $V_{\text{Lindhard}} (q=0)$ at each $r_s$ are plotted in Fig. \ref{staticomega|ee|} which shows the relative importance of exchange and correlation and the deformable background.

\begin{figure}	
    \centering
    \includegraphics[width=1.0\columnwidth]{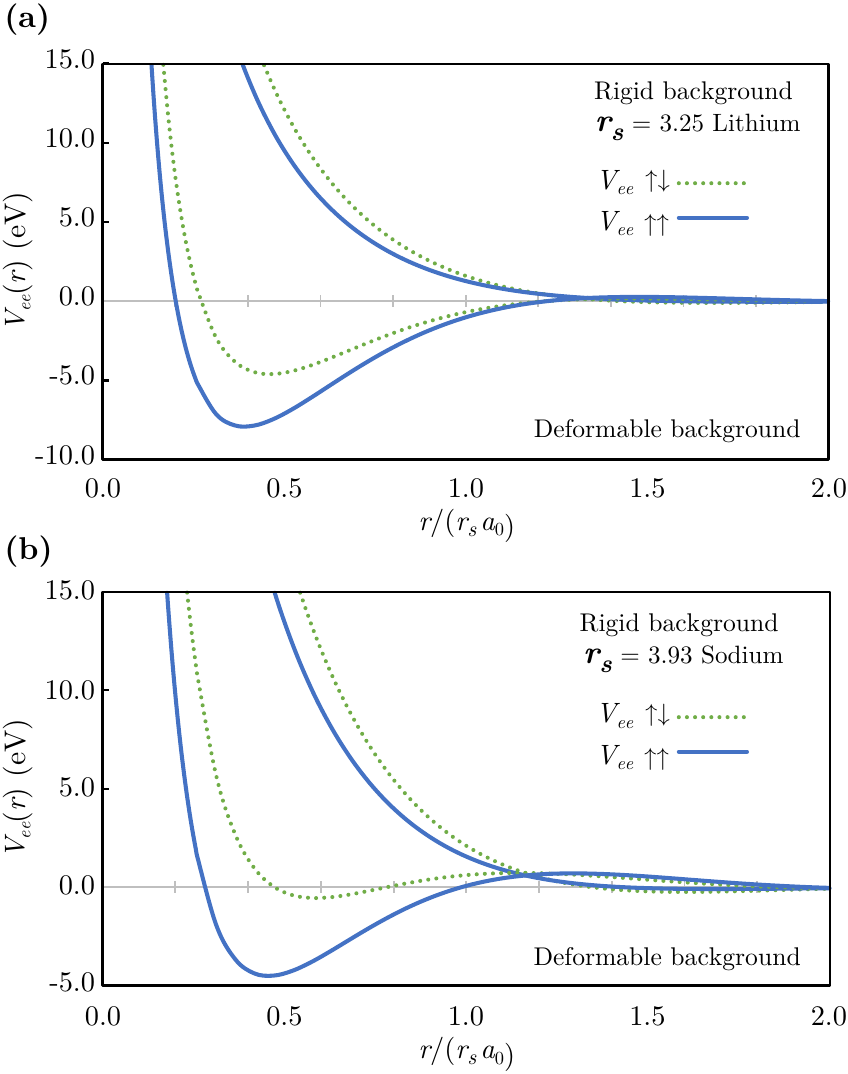}
    \caption{$V_{ee} (r)$, the Fourier transform of $V_{ee} (q)$, the electron-electron interaction shown in Fig. \ref{ee|Li,Na}, at the densities of 			lithium (a) and sodium (b). This figure compares the repulsive interaction calculated for a rigid background to the net electron-electron interaction 		when a deformable background is included.  The top two curves are for the rigid background for opposite and parallel spins. The bottom two curves 			include the deformable background contribution. }
    \label{newfig10}
	\end{figure}

Figure \ref{staticomega|ee|} shows that the effects of screening by the deformable background are as large as the effects of exchange and correlation. The $q = 0$ value of the net electron-electron for lithium is attractive for both parallel and opposite spins, while all of the other alkali metals are repulsive. Lithium is the only alkali metal that exhibits superconductivity at ambient pressure. Although the net electron-electron for lithium is more attractive for parallel spins than opposite spins, this does not necessarily imply triplet pairing because the spatial part of the overall wave function must be anti-symmetric. This is discussed briefly after Fig. \ref{newfig10}.
	
The wave vector dependent electron-electron interaction for this simple model of lattice screening in the alkali metals can be calculated using Eq. \eqref{eq13}, Eq. \eqref{Vee|phonon} and Eq. \eqref{B3}. This assumes the linear phonon dispersion relation of Eq. \eqref{eq18}. A linear phonon dispersion should be the correct behavior at small $q$, but overestimates $\omega_q$ as $q$ approaches $2\,k_F$ (and a reciprocal lattice vector). Since $\omega_q^2$ is in the denominator, a smaller value near $2 \,k_F$ would imply an even more attractive potential. The results for lithium and sodium are shown in Fig. \ref{ee|Li,Na}.
	
The electron-electron interaction in a rigid background is repulsive.  The repulsion is less for parallel spins than for opposite spins. The deformable background at low frequencies contributes a negative term due the additional screening by the deformable background/lattice. The background screening is due to the net Coulomb interaction and is independent of spin for a non-magnetized electron gas.

The most interesting feature is that the net electron-electron interaction in lithium is attractive from $q=0$ to above $2\,k_F$, for both parallel and opposite spins. A more careful treatment of the background/phonons is needed, but this result is likely to be qualitatively correct. The strong attractive region in lithium may explain the source of the observed superconductivity. 

The other alkali metals with lower densities have more repulsive electron-electron interactions and smaller effects of the deformable lattice as measured by the bulk modulus. The lattice screening is still a significant effect.  The repulsive interaction in the rigid lattice falls off more quickly with wave vector than the attractive contribution from the deformable background. The resulting net electron-electron interaction has a minimum near $1.5-2 \, k_F$. This minimum is slightly repulsive for opposite spins, and slightly attractive for parallel spins for all the alkali metals. The small attraction for parallel spins near $2 \, k_F$ may lead to interesting physics. Sodium is shown in Fig. \ref{ee|Li,Na}. 
The curves for the other alkali metals are similar.

Figure \ref{newfig10} shows $V_{ee}(r)$, the Fourier transform of the electron-electron interactions in Fig \ref{ee|Li,Na}. The energy scale is $eV$, and the Fermi energy for lithium is $4.74 \, eV$ and for sodium it is $3.23 \,  eV$. With a rigid lattice, the electron-electron interaction is repulsive as shown in the top two curves of Figure \ref{newfig10}. The classical turning points for scattering, where the repulsive potential is equal to the Fermi energy, are approximately $0.75 \, r_s$ for lithium and $0.9 r_s$ for sodium. Electron-electron scattering does not contribute to the electrical resistivity because momentum and charge are conserved in the scattering event (except for a small effect due to umklapp scattering in a real-lattice as opposed to a uniform background). Electron-electron scattering does however contribute to the thermal resistivity.

When the deformable background is included, the overall interaction potential for lithium is attractive for both parallel and opposite spins in the region $0.25-1.0 \, r_s$ with a depth comparable to the Fermi energy, with a repulsive core at shorter distances. At larger distances, not shown in Fig \ref{newfig10}, much smaller oscillations similar to Friedel oscillations are seen. The interaction or scattering of two electrons in the electron gas with an attractive potential due to the deformable background could include resonances, virtual bound states or even bound states. We note that although the electron-electron attraction is stronger for parallel spins, the overall wave function which must be anti-symmetric would imply that the spatial part is anti-symmetric (p-wave) and the probability that the electron is found at small distances is lower. Thus it is likely that the opposite spin electrons with a symmetric (s-wave) wave function would sample more of the region of the attractive potential. 

For sodium, only the electron-electron interaction for parallel spins becomes significantly attractive.

We present this Fourier transform real space calculation to stimulate thinking within the single particle picture of the electron gas-- which has limitations. The usual thinking is in momentum and frequency space.  Equation \eqref{eq13} gives the explicit frequency dependence. The local field factors used here are independent of frequency, but the frequency dependence of the Lindhard function is known.

The electron-electron interaction in this paper can be used in calculations of superconductivity. The simple inclusion of the deformable lattice shows that lithium is substantially different from the other alkali metals, and lithium is the only alkali that shows superconductivity at ambient pressure. Cesium exhibits superconductivity at high pressure. The net electron-electron interaction is sensitive to exchange and correlation and to the deformable lattice. Richardson and Ashcroft\cite{ref9} considered superconductivity with the KO electron-electron interaction and phonons treated on an equal basis, and applied the theory to several metals including lithium. The considerable difficulties in comparison with experiment are discussed in that paper.

The region of largest static attraction between two electrons in lithium is at very short distances. The role of this short range attraction in the dynamical theory of superconductivity is not known to us.

A strong word of caution is needed when comparing electron gas calculations with experiment, particularly at low density close to the compressibility divergence at $r_s=5.25$. Factors such as effective mass, core polarization and renormalization factor $z$ have to be carefully considered when comparing with experiment.  These effects were discussed in Kukkonen and Wilkins\cite{ref6} and in Ref\cite{ref9}. A simple example is Cesium with $r_s=5.62$ which is beyond the compressibility instability. Cesium and the other alkali metals have polarizable cores with a core polarization dielectric function $\epsilon_B$ that is frequency independent in the regions of interest. The correct theory\cite{ref6} is an electron gas with electrons of effective mass $m^\ast$ in a neutralizing uniform positive background with dielectric constant $\epsilon_B$. The result is obtained by scaling the known solutions for the electron gas with mass $m$ and a non-polarizable background, but at a different density $r_s^\ast = r_s (m^\ast /m \epsilon_B)$ and evaluated at a scaled wave vector. For cesium, $\epsilon_B= 1.27$ and this re-normalizes $r_s$ from 5.62 to 4.44 which is below the compressibility divergence. The core polarization corrections for lithium and sodium are quite small.

\section{Summary}
Variational Diagrammatic Monte Carlo (VDMC) calculations of the wave vector dependent spin local field factor (exchange and correlation kernel) have been presented and utilized for all calculations. Using the density and spin local field factors and explicit equations presented in this paper, all of the response functions of the three-dimensional electron gas can be easily and quantitatively calculated.  Exchange and correlation are fully included within the self-consistent local field approximation and the compressibility and susceptibility sum rules at $q=0$ are satisfied.

The full spin dependent electron-electron interaction is calculated using these field factors. For a rigid background, the electron-electron interaction shows no divergence at the compressibility divergence, is repulsive for all $r_s$ in the metallic region.

Considering a deformable background, the $\omega=0$ (static) screening by the background is shown to be very important with an effect that is as large as exchange and correlation. A simple calculation shows that with a deformable background modelled by using the measured bulk modulus, the net electron-electron interaction including exchange and correlation is attractive (negative) in a large range of momentum and real space for lithium which does exhibit superconductivity at ambient pressure, and is mostly positive (repulsive) for all the other alkali metals.

The compressibility divergence does not appear in the electron-electron and electron-test charge interactions, but still shows up in the test charge-test charge interaction as a divergence in the dielectric function and in the screening by the deformable background.

The quantitative spin dependent electron-electron interaction can be used in other calculations such as superconductivity, and can be also used as a starting point for improved numerical calculations.  The self-consistent local field calculation of the KO interaction using Feynman diagrams may lead to new techniques for identifying massive cancellations of divergent diagrams and allow new perturbation techniques around the self-consistent solution.

More detailed numerical calculations of the density local field factor $G_+(q)$ from $q=0$ to $3 \, k_F$ would be welcomed, as well as a physical explanation for the behavior.

\section*{Acknowledgements}
The authors are grateful to William Halperin, Jan Herbst, Giovanni Vignale and David Ceperley for discussions, and to Giovanni Onida and Massimiliano Corradini for providing the exact coefficients for their analytic function for $G_+(q)$ that fits the QMC data, and to Tori Hagiya for comparing his experimental data with our theory.

C.A.K. acknowledges support from the US Social Security Administration. K.C. appreciates support from the Simons Foundation. 

\appendix
\section{DENSITY LOCAL FIELD FACTOR    $\mathbf{G_+(q)}$}
The density local field factor has been a subject of research for more than 60 years. $G_+(q)$  is needed to calculate the dielectric function and vertex correction. These quantities are sufficient to calculate all of the interactions and response functions that do not depend on spin. Many proposals have been made for the dielectric function and $G_+(q)$ particularly by Hubbard\cite{ref10}, Geldart\cite{ref11} and collaborators, and Singwi\cite{ref12} and collaborators. When it was realized that the compressibility sum rule was of paramount importance at $q=0$, the dielectric functions and $G$’s were manually adjusted to satisfy the sum rule even though the calculations themselves did not satisfy the sum rule. The calculations were used to interpolate the wave vector dependence above $q=0$. The behavior of $G_+(q)$ at $2 \, k_F$  and above has been the subject of considerable research and debate, but is not important for the calculations of the static response properties of the electron gas because the Lindhard  function cuts off the response quickly above $2 \, k_F$ .

The Quantum Monte Carlo method was used by Moroni, Ceperley and Senatore\cite{ref13} to calculate $G_+(q)$ from $q=k_F$  to $q=4 \, k_F$  for $r_s = 2, 5$ and $10$. Corradini, Del Sole, Onida and Palummo\cite{ref14} fitted the QMC data with an analytical function that reflected the appropriate small and large limits. This expression is given in terms of derivatives of the electron gas energy. The Quantum Monte Carlo calculations are considered the most accurate, but there are no data below $q=k_F$.  Richardson and Ashcroft\cite{ref15} calculated the local field corrections at finite frequencies, and their static results are similar, but differ somewhat in detail from the QMC results. They emphasized the importance of the sum rules. Richardson and Ashcroft also provided an interpolation formula. Retrospective discussions of the local field factors were given by Simion and Giuliani\cite{ref16} and by Hellal, Gasser and Issolah\cite{ref17}.  

\begin{figure}		
    \centering
    \figuretitle{(a)}
    \includegraphics[width=1.0\columnwidth]{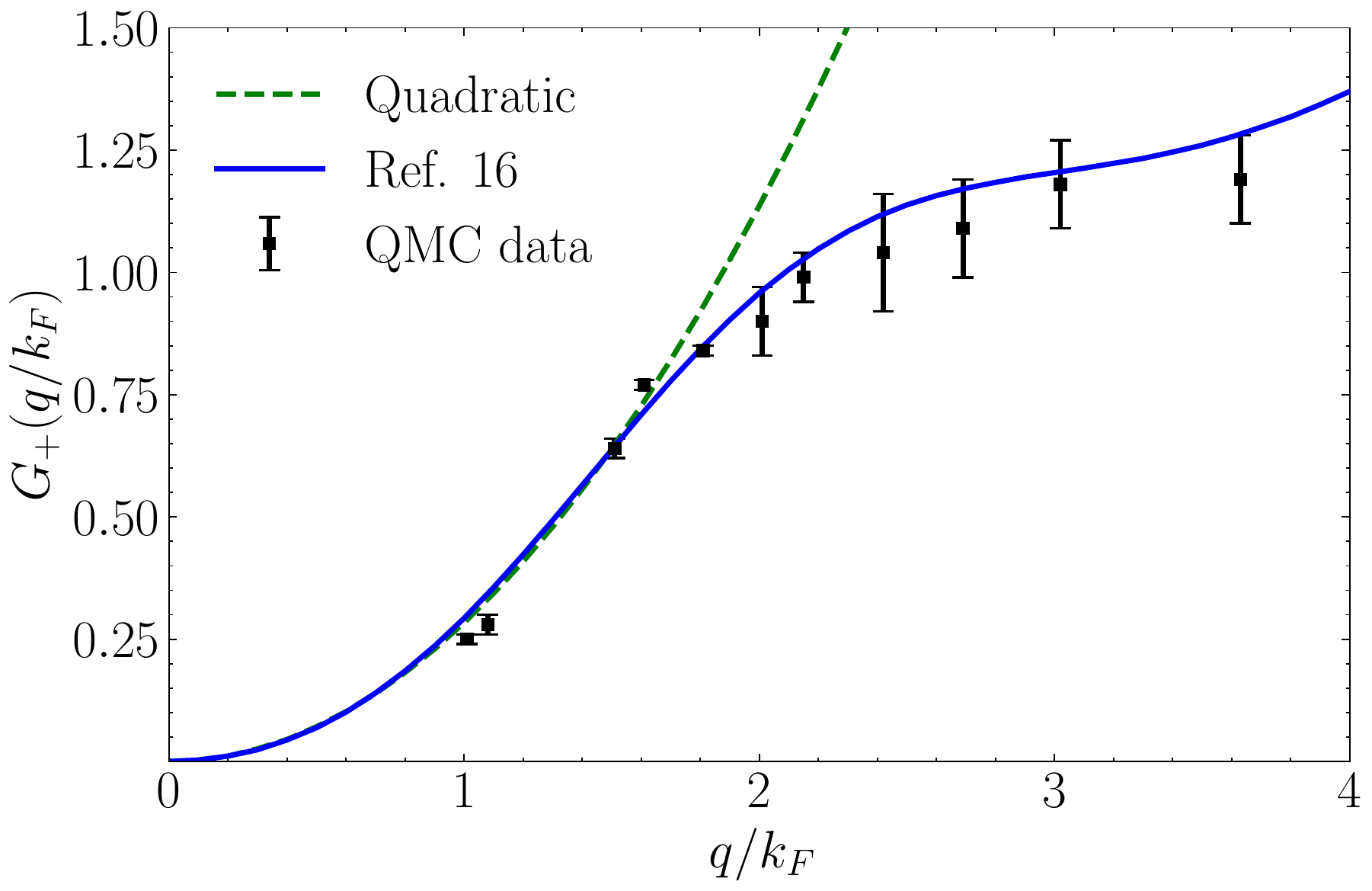}
    \figuretitle{(b)}
    \includegraphics[width=1.0\columnwidth]{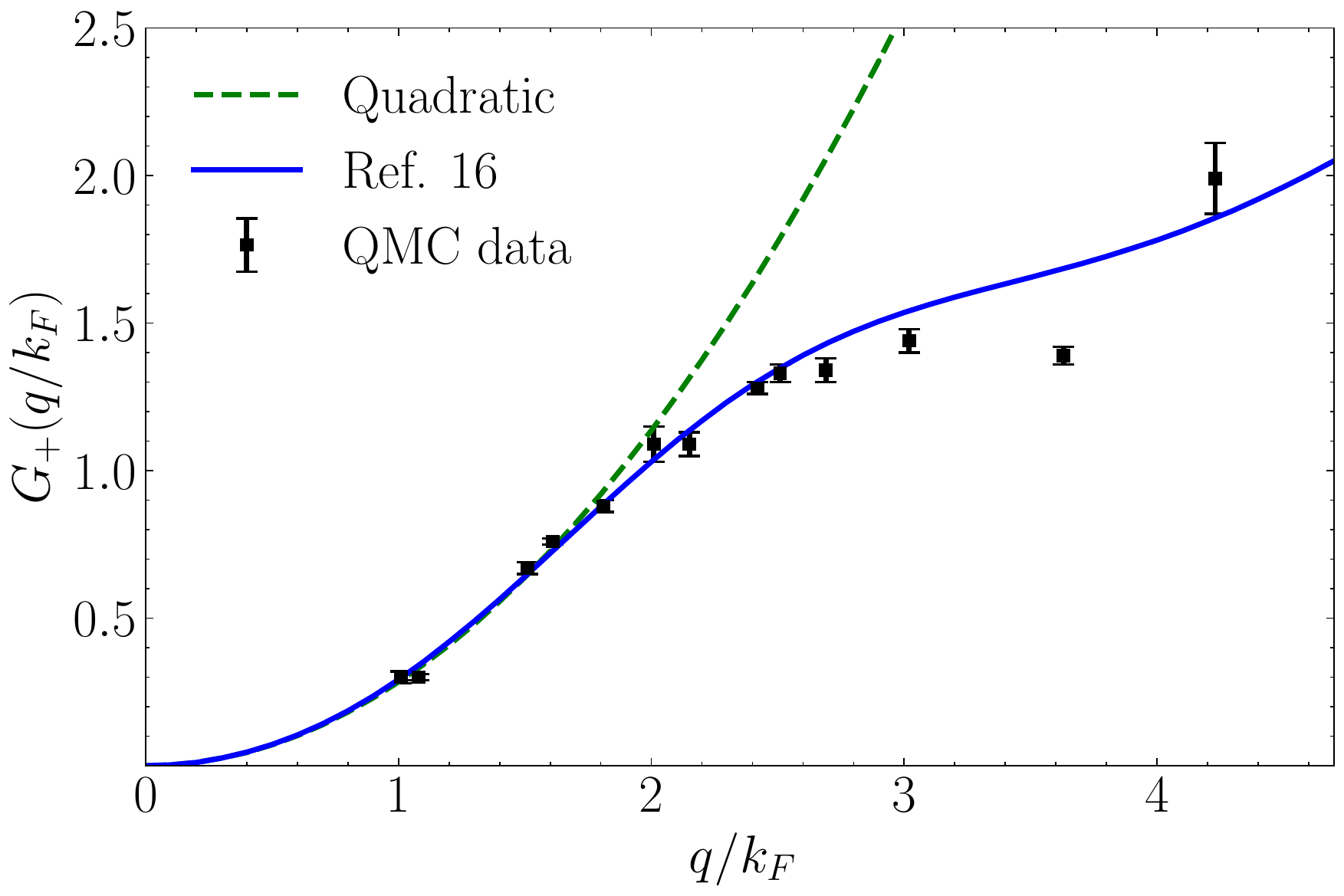}
    \caption{Density local field function $G_+(q)$ plotted versus $q/k_F$  for $r_s=2$ (a) and $r_s=5$ (b).  Data points are the Quantum Monte Carlo 				calculations from Ref. \cite{ref10}. The solid curves that fit the data are the analytic function from Ref. \cite{ref11}. The quadratic Eq. \eqref{eq4} is 			the proposed simple approximation to $G_+(q)$ for calculating the response functions. Error bars are shown for all data points. If they are not 			evident, the error is smaller than the data point.}
    \label{densitylocalfieldG+}
	\end{figure}

We plot the Quantum Monte Carlo results of Ref. \cite{ref10} for $G_+(q)$ together with the analytic interpolation function Ref.\cite{ref11}, and the even simpler quadratic formula that quite accurately reproduces the response functions of the electron gas in Fig. \ref{densitylocalfieldG+}. 
Although there are no data below $q=k_F$ , the QMC results above $k_F$ show that $G_+(q)$ follows the quadratic required by the compressibility sum rule up until nearly $2 \, k_F$  and then falls significantly below the initial quadratic. Theory predicts that the large $q$ behavior will also be a quadratic but with a different coefficient.  Other earlier versions of $G_+(q)$ \cite{ref10} had a much smaller value at $2 \, k_F$.  
  

Figure \ref{densitylocalfieldG+} emphasizes the large $q$ behavior of $G_+(q)$. However the static response functions of the electron gas depend mostly on the low $q$ behavior, because 
the Lindhard functions cuts off the effect of $G_+(q)$ quickly above $q=2 \, k_F$ .

At small $q$, the compressibility sum rule specifies that
\begin{eqnarray}\label{G+to0}
	G_+(q \rightarrow 0) = \left( 1-\frac{\kappa_0}{\kappa}	\right)\left(\frac{q}{q_{TF}}\right)^2
\end{eqnarray}

To emphasize the low q behavior, $G_+(q)/(q/q_{TF})^2$ is plotted in Fig. \ref{densitylocalfieldG+/qTF} below, where the intercept at $q=0$ is $(1-\kappa_0/\kappa)$. Note that the density exchange and correlation kernel needed for Time-Dependent Density Functional Theory is given by $f_{xc}= -4\pi G_+(q)/(q/q_{TF})^2$.

	\begin{figure}[h!]		
    \centering
    \figuretitle{(a)}
    \includegraphics[width=1.0\columnwidth]{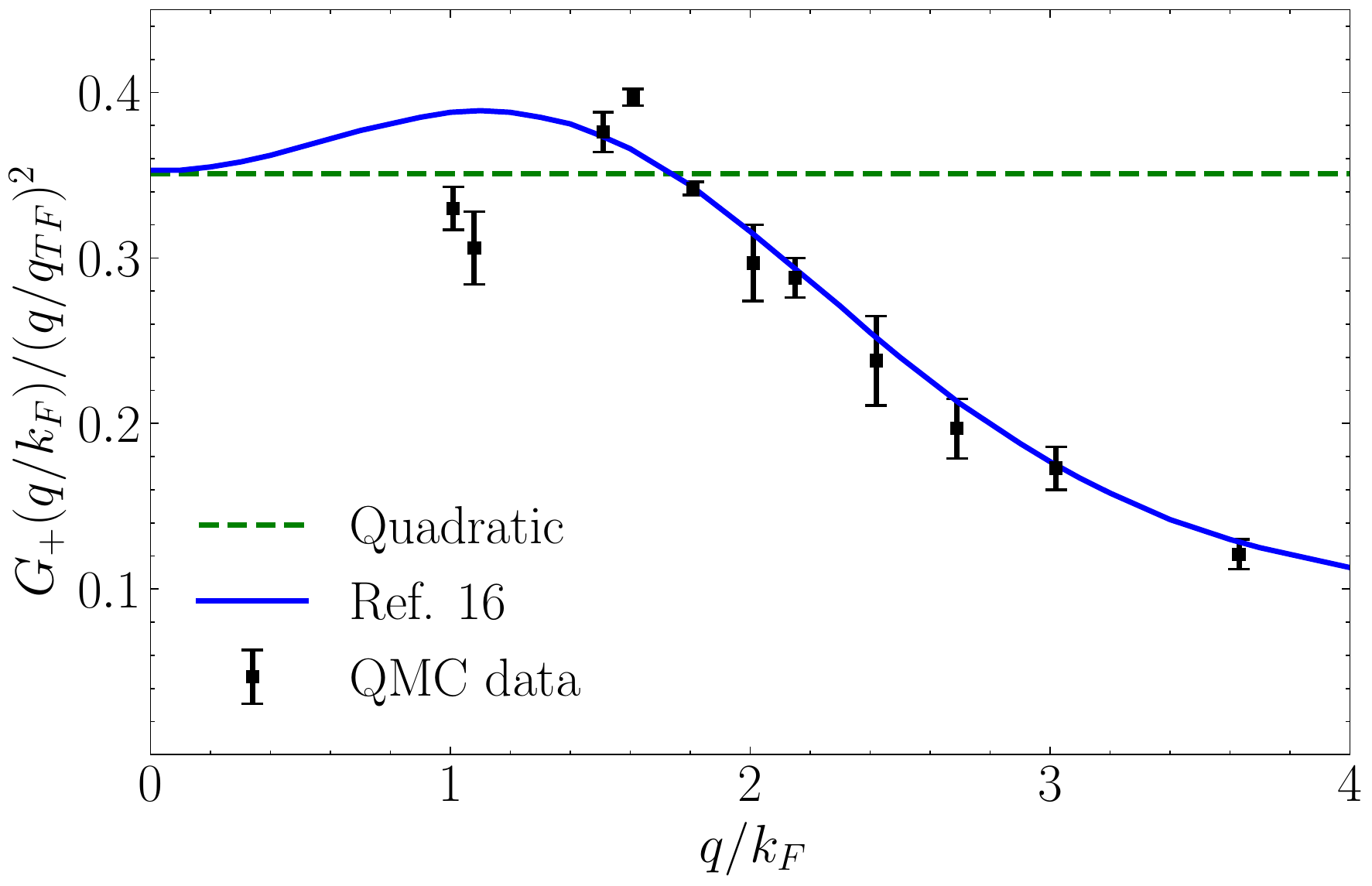}
    \figuretitle{(b)}
    \includegraphics[width=1.0\columnwidth]{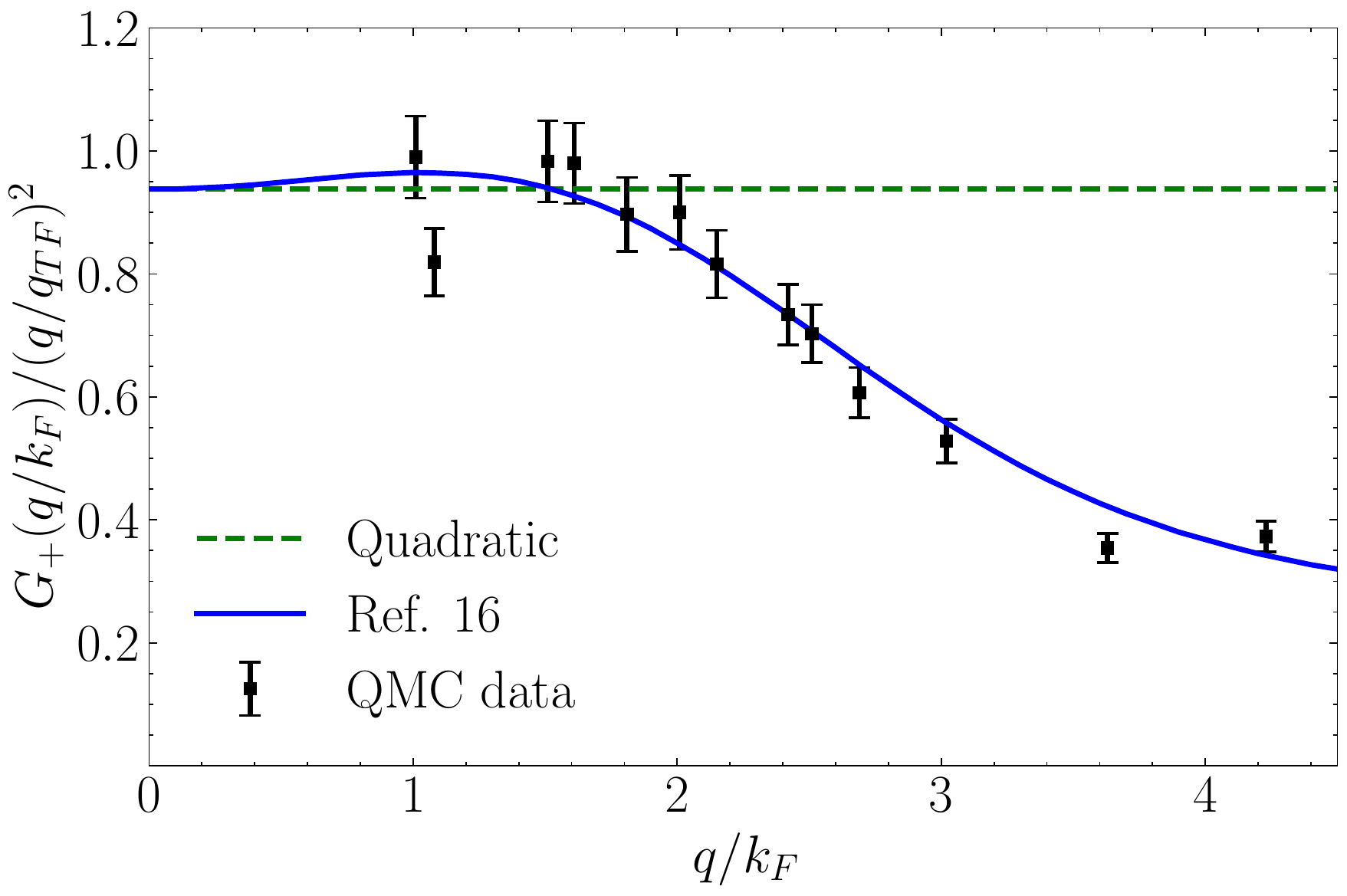}
    \caption{$G_+(q)/(q/q_{TF})^2$, the density local field function divided by $(q/q_{TF})^2$ is plotted versus $q/k_F$ for $r_s=2$ (a) and $r_s=5$ (b). Data points are the Quantum Monte Carlo calculations from Ref. \cite{ref13}. The solid curves that fit the data are the analytic function from Ref. \cite{ref14}. The straight line represents the quadratic Eq. \eqref{eq4} that is the proposed simple approximation to $G_+(q)$ for calculating the 				response functions. Error bars are shown for all data points. }
    \label{densitylocalfieldG+/qTF}
	\end{figure}

\begin{figure}[h!]		
    \centering
    \figuretitle{(a)}
    \includegraphics[width=1.0\columnwidth]{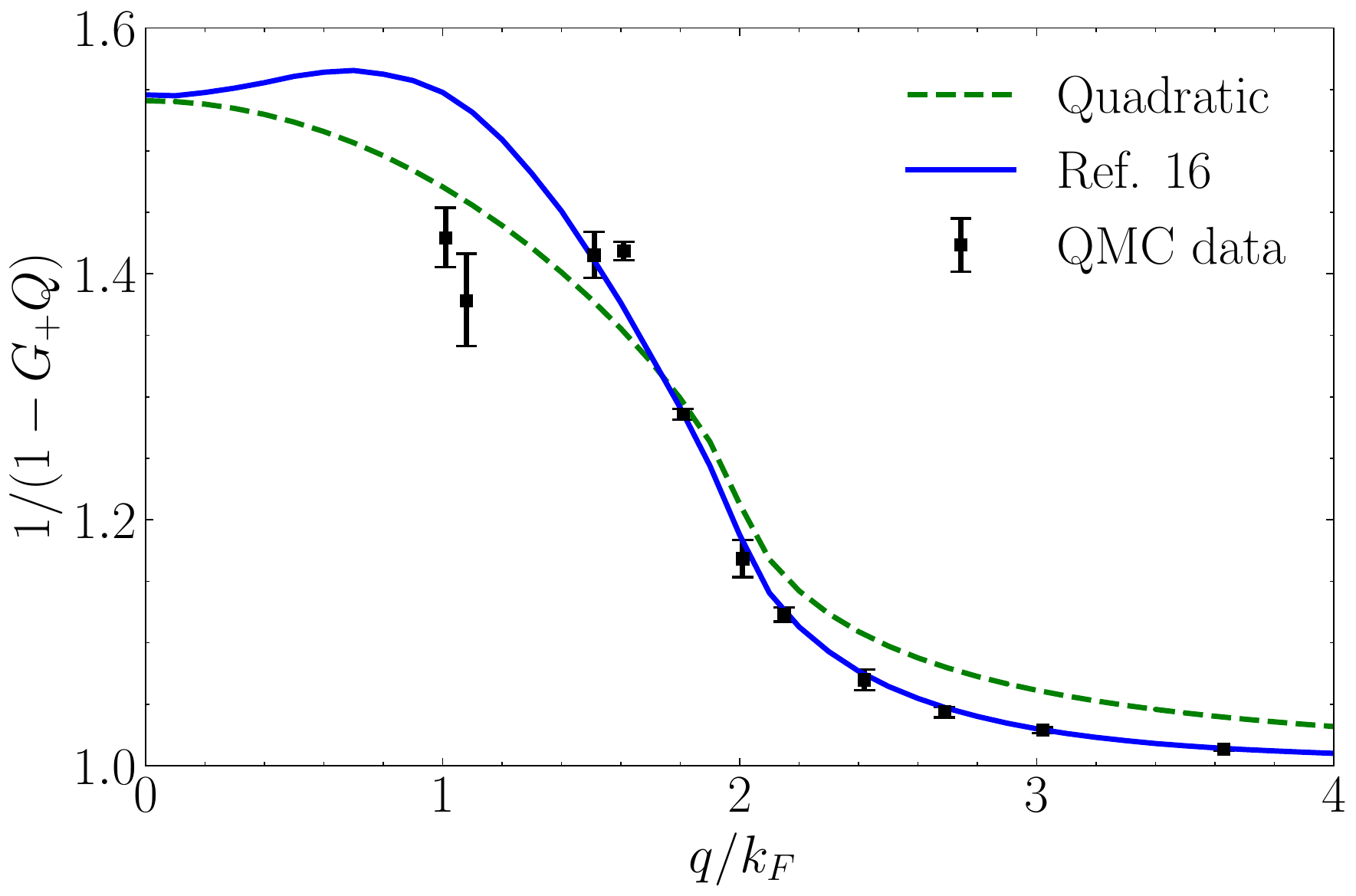}
    \figuretitle{(b)}
    \includegraphics[width=1.0\columnwidth]{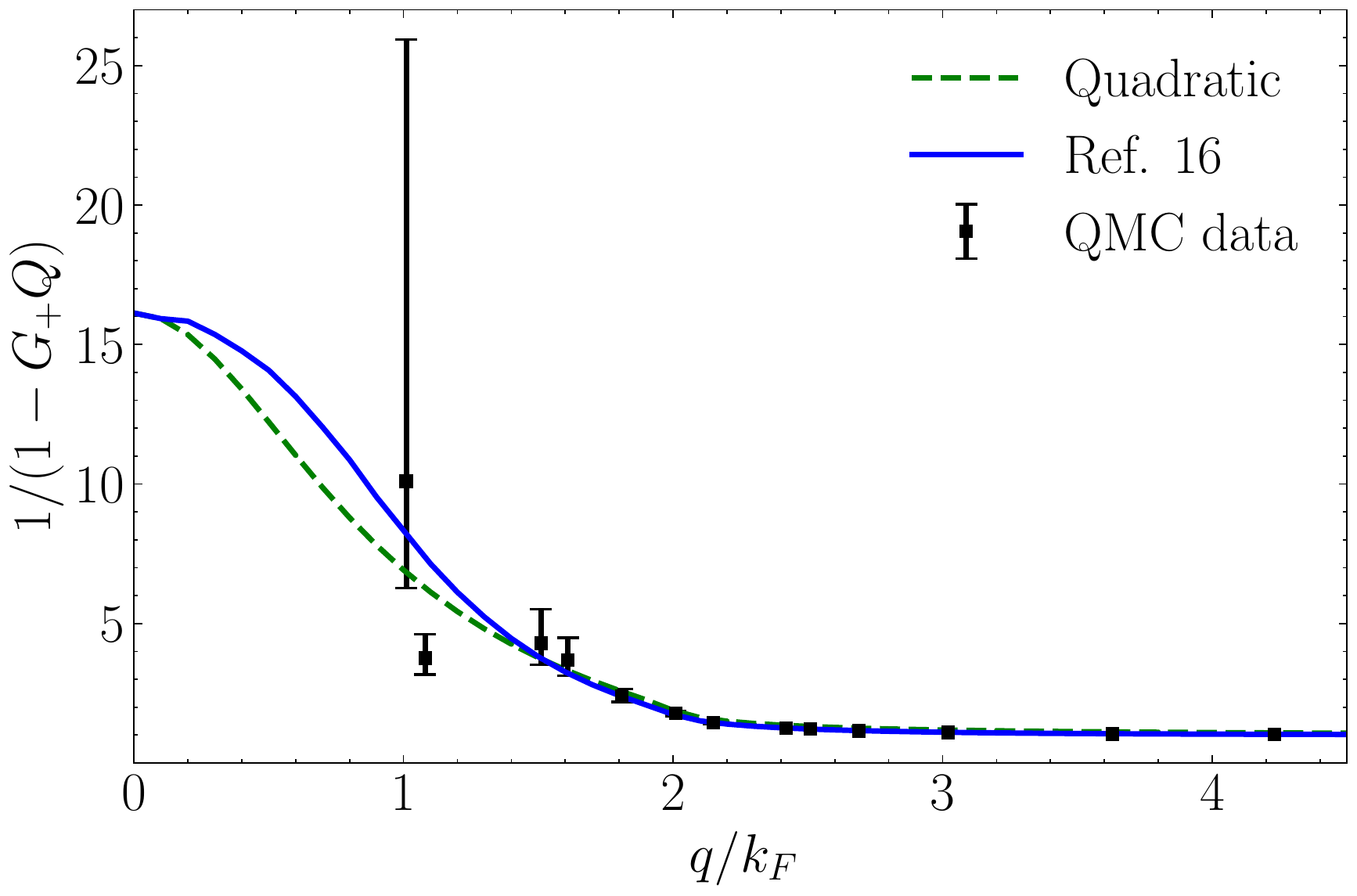}
    \caption{Vertex function $\Lambda = 1/(1-G_+Q)$ plotted versus $q/k_F$ for $r_s=2$ (a) and $r_s=5$ (b). Data points are the Quantum Monte Carlo 			calculations from Ref. \cite{ref13}. The dotted curve that fits the data is the analytic function from Ref. \cite{ref14}. The solid curve uses Eq. \eqref{eq4}, the proposed simple quadratic approximation for calculating the response functions. Error bars are shown for all data points. If they are 			not evident, the error is smaller than the data point. Note that the $y-\text{axis}$ starts at 1.0 for $r_s=2$, in order to emphasize small 				differences. }
    \label{Lambdavsq/kf}
	\end{figure}

Note again that there are no Quantum Monte Carlo data below $q=k_F$.  The $q=0$ values of the quadratic and the analytic function are set by the compressibility sum rule. The compressibility is also calculated by Monte Carlo methods, and the $q=0$ value is much more accurate than the q dependent data. The compressibility sum rule dictates $G_+(0)$ and that the initial behavior will be quadratic, Eq. \eqref{G+to0}, as represented by the constant horizontal line. The analytic interpolation formula \eqref{14} apparently fits the data above $1.5 k_F$ quite well, but misses substantially the data at $k_F$ for $r_s=2$. This illustrates the problem of global curve fitting with analytical functions with a limited number of parameters. The simple quadratic does at least as good a job below $2k_F$ and is substantially different at larger $q$, but the effect at large $q$ is not very important for the response functions as will be shown below. Looking carefully at the QMC data between $k_F$ and $2 \, k_F$, there is an intriguing hint of structure in $G_+(q)$, with data both below and above the quadratic. Further Quantum Monte Carlo calculations from $q=0-3 \,k_F$ would be informative.  
The vertex function which embodies the effect of exchange and correlation in the response functions is plotted in Fig \ref{Lambdavsq/kf}, where the simple quadratic is compared to the QMC data and the fitting function for this data.

The vertex function $\Lambda = 1 /(1-G_+Q)$ is required to calculate the dielectric function, and the other interactions in the electron gas. Figure \ref{Lambdavsq/kf} shows the vertex function using the actual QMC data and the analytic function as well as the simple quadratic.  The first point to note is that the $q=0$ value of the vertex function is exactly given by the compressibility sum rule and is equal to $\Lambda(0) = \kappa/\kappa_0$  which diverges at the compressibility divergence as approximately $1/(1-r_s/5.25)$. At $r_s=2$, the vertex enhancement is substantial, but it is not near the compressibility divergence. At $r_s=5$, the electron gas is very near the divergence and the vertex enhancement is very large. The error bar in the value of $G_+(q)$ at $q=k_F$ reaches nearly a point of instability and its effect is magnified because it appears in the denominator. The vertex function is extremely sensitive to small changes at $r_s=5$.

The simple quadratic function for $G_+(q)$  yields a vertex function that satisfies the compressibility sum rule and fits the vertex function derived from the QMC data as well as the fitting formula of Ref. \cite{ref11}. This is despite the large differences at large $q$, because the contributions at large $q$ are cut off by the Lindhard function.  The fitting function should be used for any calculations that depend on $q$ substantially above $2k_F$.

The fact that the vertex function resulting from the analytic function is larger than the quadratic for $q$ greater than zero and less than $k_F$ is entirely due to the curve fitting, and there are no QMC data in this region. 

The simple quadratic approximation for $G_+(q)$ was suggested by Taylor \cite{ref18} 40 years ago and we concur.

\begin{figure}[h!]		
    \centering
    \includegraphics[width=1.0\columnwidth]{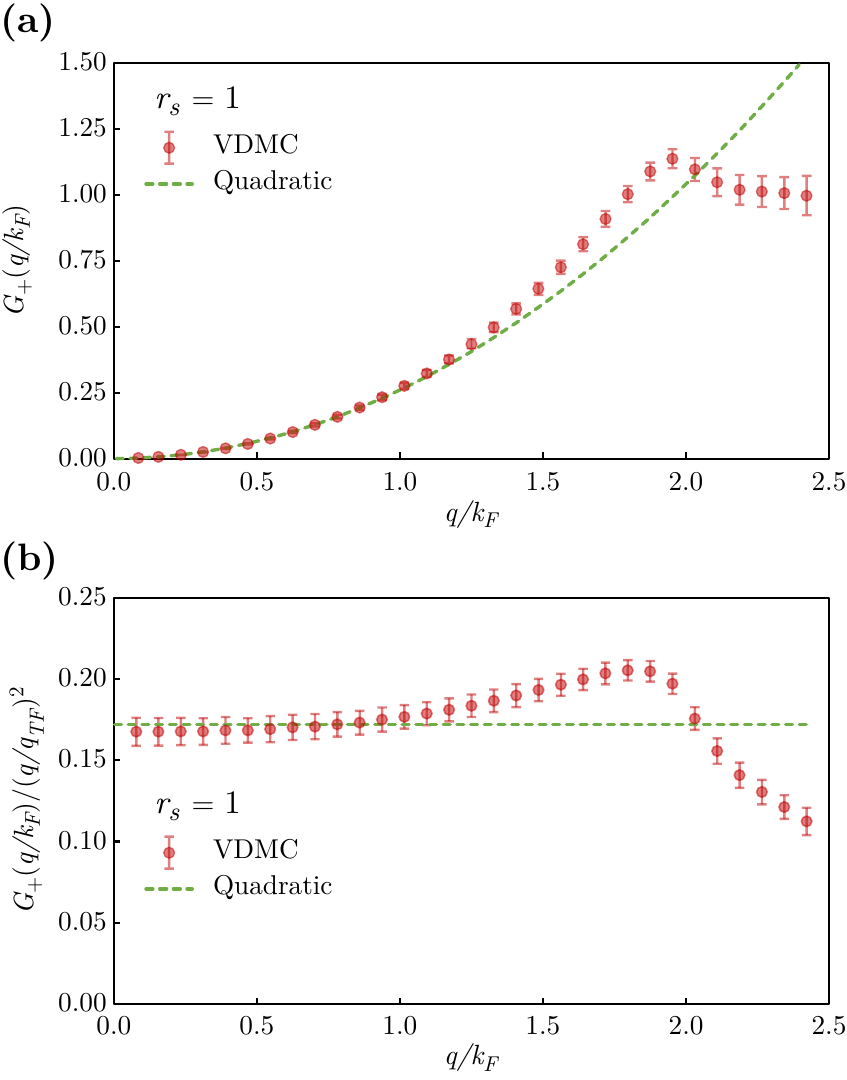}
    \caption{Density local field factor $G_+(q)$ from VDMC calculations for $r_s =1$ plotted versus $q/k_F$ (a), and $G_+(q)/(q/q_{TF})^2$ (b).  Error bars are shown. }
    \label{catorce-fourteen}
	\end{figure}
	
\begin{figure}[h!]		
    \centering
    \includegraphics[width=1.0\columnwidth]{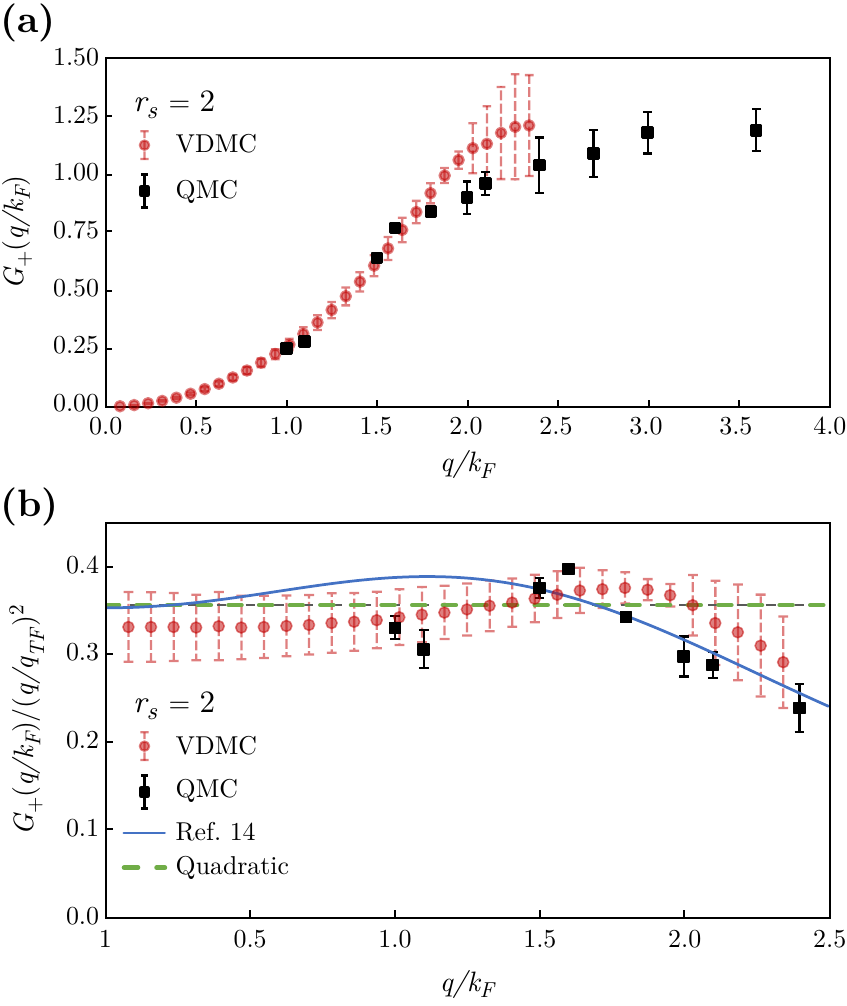}
    \caption{Density local field factor $G_+(q)$ from VDMC calculations for $r_s =1$ plotted versus $q/k_F$ (a), and $G_+(q)/(q/q_{TF})^2$ (b).  Error bars are shown. }
    \label{quince-fifteen}
	\end{figure}

The current VDMC method works well when the vertex correction or susceptibility enhancement is modest.  This corresponds to 				$\chi_0/\chi$ or $\kappa_0/\kappa<0.6$ in Figure \ref{kappa}. For the spin susceptibility, this is near $r_s=5$.  For the compressibility, this corresponds to $r_s = 2$. As a test, we have calculated the density local field factor $G_+(q)$ for $r_s = 1$ and $2$, and the data are shown in Figures \ref{catorce-fourteen} and \ref{quince-fifteen}. At $r_s = 2$, we compare the new VDMC calculations to QMC results of Ref. \cite{ref13} and the corresponding interpolation formula of Ref. \cite{ref14}.
	
	At $r_s = 1$, Fig. \ref{catorce-fourteen} shows that the density local field factor $G_+(q)$ has the same qualitative behavior as the spin local field factor $G_-(q)$ which are shown in Fig. \ref{Gminus} and Fig. \ref{ratio}. The error bars are acceptable. Both show that $G$ rises above the quadratic at approximately $1.2 \,  kF$ and then falls below at $2 \, kF$.

	The new data for $G_+(q)$ in Fig. \ref{quince-fifteen} demonstrates the limitations of the current version of the VDMC approach. Ordinarily, we would not show data with such large error bars. However, we want to compare with the QMC data Ref. \cite{ref13} and to provide new data below $k_F$. The VDMC data with error bars overlap the QMC data with its error bars except for two points near $2\, k_F$, and even there, the agreement is quite close. The simple quadratic approximation for $G_+(q)$ given in Eq. \eqref{eq4} represents the VDMC data quite well up to $2\, k_F$.  The interpolation curve of Ref. \cite{ref14} fits the data well above $2\, k_F$. The data for $G_+(q)$ at $r_s =2$ are qualitatively similar to the data for $G_-(q)$.
	
	The VDMC data for $G_-$ and the limited data for $G_+$ show that both of these local field factors are smooth functions of wave vector. Although the data rise slightly above the quadratic between 1.5 and $2 \, k_F$, there is no evidence of a large peak.  We have not developed a physical intuition for this behavior.
	
\section{TEST CHARGE-TEST CHARGE $\mathbf{V_{tt}}$ AND ELECTRON-TEST CHARGE $\mathbf{V_{et}}$ INTERACTIONS AND DENSITY RESPONSE FUNCTION}
We use the same quadratic function for $G_+(q)$ to plot the test charge test-charge and electron-test charge interactions at $r_s=2$ and $5$, and compare them to the Lindhard interaction. For the general reader, a simple physically motivated derivation of these interactions and the electron-electron interaction are in Ref. \cite{ref1}.

The test charge-test charge interaction $V_{tt}$ is the Coulomb potential generated by a test charge plus the induced screening cloud, and felt by another test charge.
	\begin{figure}[h!]	
    \centering
    \includegraphics[width=1.0\columnwidth]{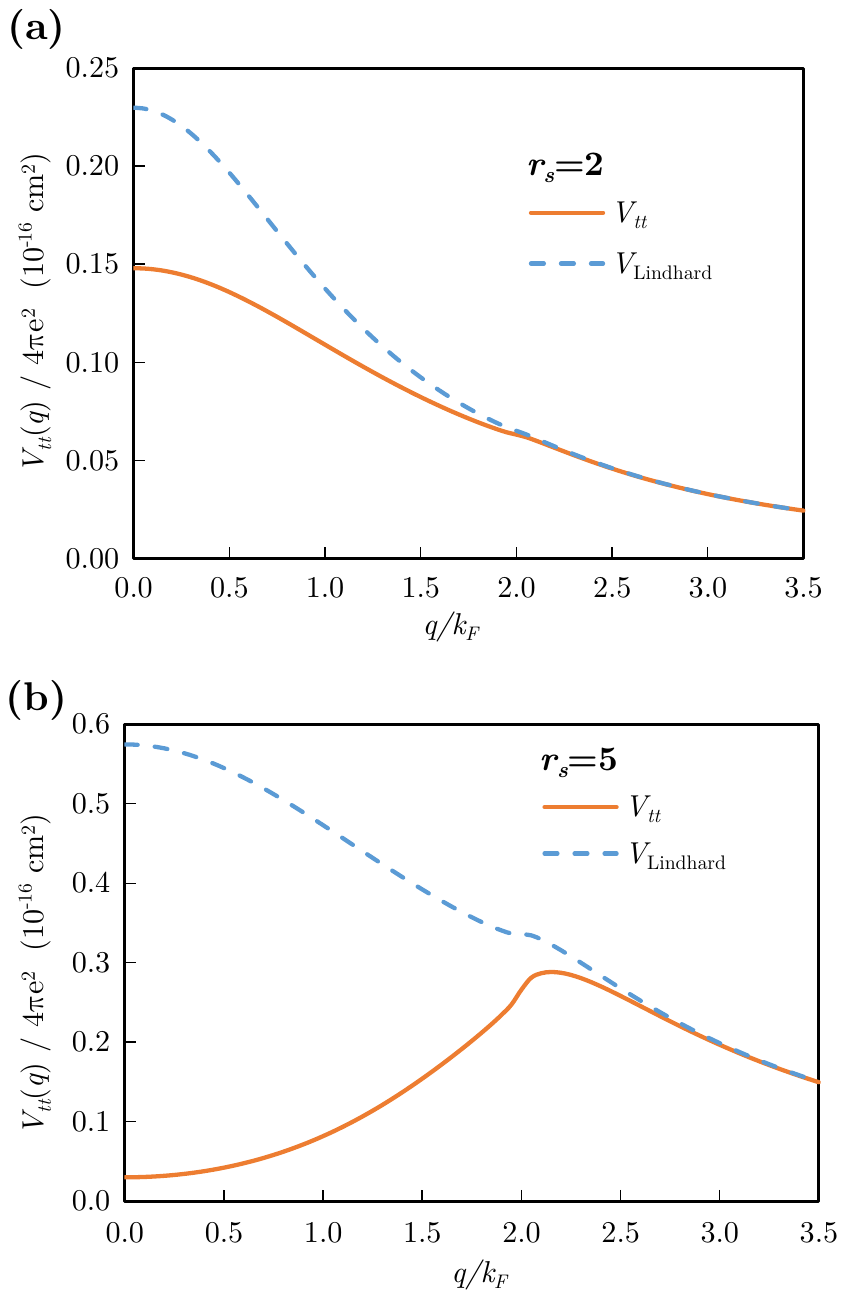}
    \caption{$V_{tt}(q)$, the test charge-test charge or Coulomb interaction at $r_s=2$ and $5$. The potential is measured in units of $4\pi e^2$. The 		dashed line is the Lindhard potential which is equal to the Thomas Fermi potential at $q=0$ with the value $V_{\text{Lindhard}}(0) = 1/q_{TF}^2$. }
    \label{VttVlindhardvsq/kf}
	\end{figure}
	
The dielectric function $\epsilon(q,\omega)$ is defined by
\begin{equation}
    V_{tt} = \frac{V_{ext}}{\epsilon}
\end{equation}
and is written as
\begin{equation}
    \epsilon = 1 + \frac{v \Pi^0}{1-G_+ v\Pi^0} = 1 + \Lambda Q,
\end{equation}
where $\Pi^0$ is the Lindhard free electron response function and $v=4\pi e^2/q^2$. Note that others may define the response function with a minus sign. For convenience, we define $Q = v \Pi_0$ and the vertex correction $\Lambda = 1/(1-G_+ Q)$.  The potentials are measured in units of $4 \pi e^2$ so that the Thomas Fermi and Lindhard potentials at $q=0$ are simply $1/q_{TF}^2$. Without exchange and correlation, $G_+=0$ and therefore $\Lambda=1$ and the Lindhard result is obtained. 

The electron-test charge interaction $V_{et}$ is simply the test charge - test charge interaction multiplied by the vertex function. 
\begin{equation}\label{B3}
    V_{et}= \Lambda \, V_{tt} = \frac{V_{ext}}{1+(1-G_+) Q}
\end{equation}

For these interactions and response functions of the electron gas in the metallic region, the large $q$ behavior of the local field factor is not of significant importance in most applications because the Lindhard function cuts off quickly above $q=2 \, k_F$.
	\begin{figure}	
    \includegraphics[width=1.0\columnwidth]{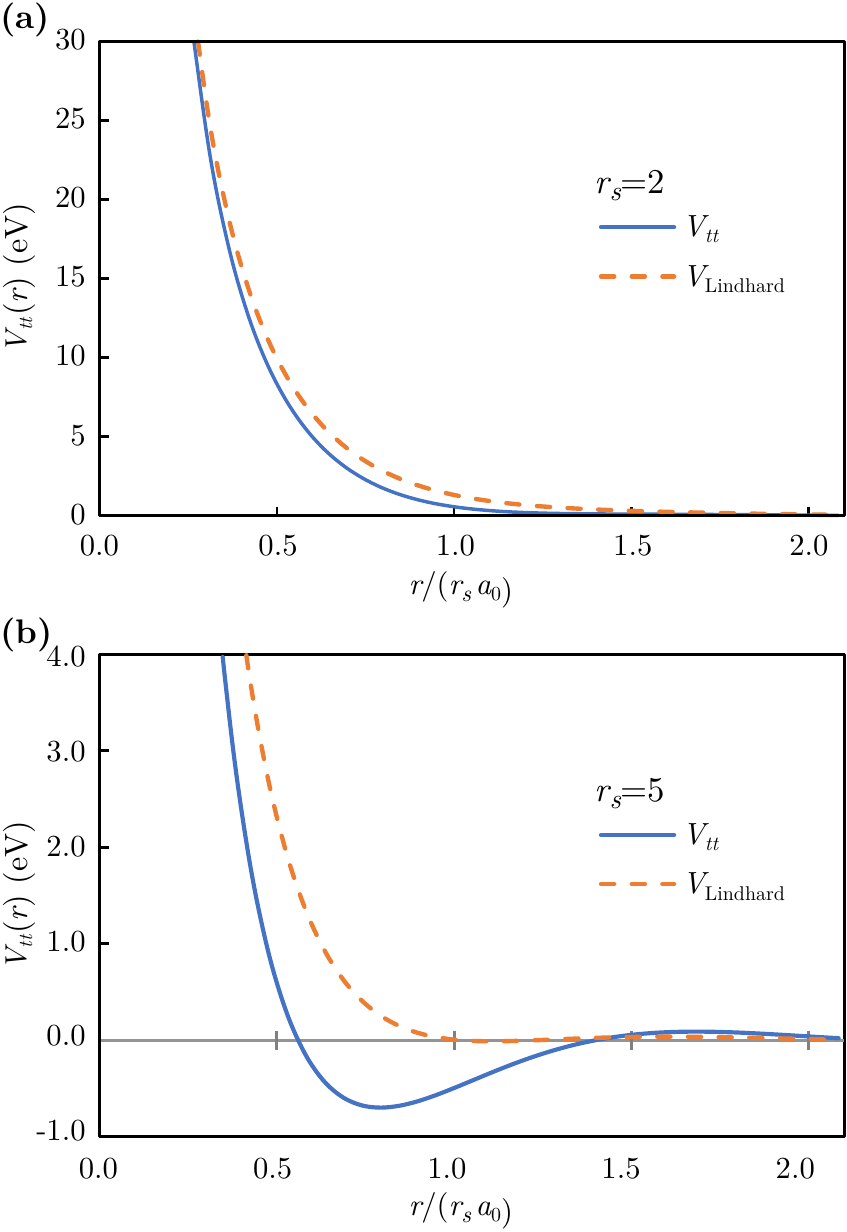}
	\caption{The test charge-test charge interactions for $r_s=2$ (a) and 5(b) have been numerically Fourier transformed and are plotted versus distance. 			The Fourier transform of the Lindhard function is shown for comparison. The data are shown from $r/(r_s \, a_0) = 0.5-2.2$. }
    \label{tc-tc vs r/()|14}
	\end{figure} 
	
The $q=0$ value of the interaction is set by the compressibility sum rule.  $V_{tt} (0) = (\kappa_0/\kappa)/q_{TF}^2$ is always less than the Lindhard or Thomas Fermi value. 
Both are virtually identical above $2.5 \, k_F$. The $q$ dependence of $G_+ (q)$ is only important between $0$ and $2 \, k_F$.

$V_{tt} (q)$ in Fig \ref{VttVlindhardvsq/kf} looks qualitatively like the Lindhard potential at $r_s=2$, but is dramatically different at $r_s\!=\!5$. This is due to the compressibility sum rule which fixes the value at $q=0$, and $G_+ (q)$ interpolates between $q=0$ and $2 \, k_F$. According to the compressibility sum rule, $V_{tt} (q=0) = 0$ at the compressibility divergence at $r_s = 5.25$, and becomes negative at larger $r_s$. It must turn positive again and match $1/q^2$ at $q$ beyond $2 \,  kF$.  We don’t have a physical intuition for this``overscreening" behavior resulting from a negative dielectric function.
\begin{figure} 	
 	\includegraphics[width=1.0\columnwidth]{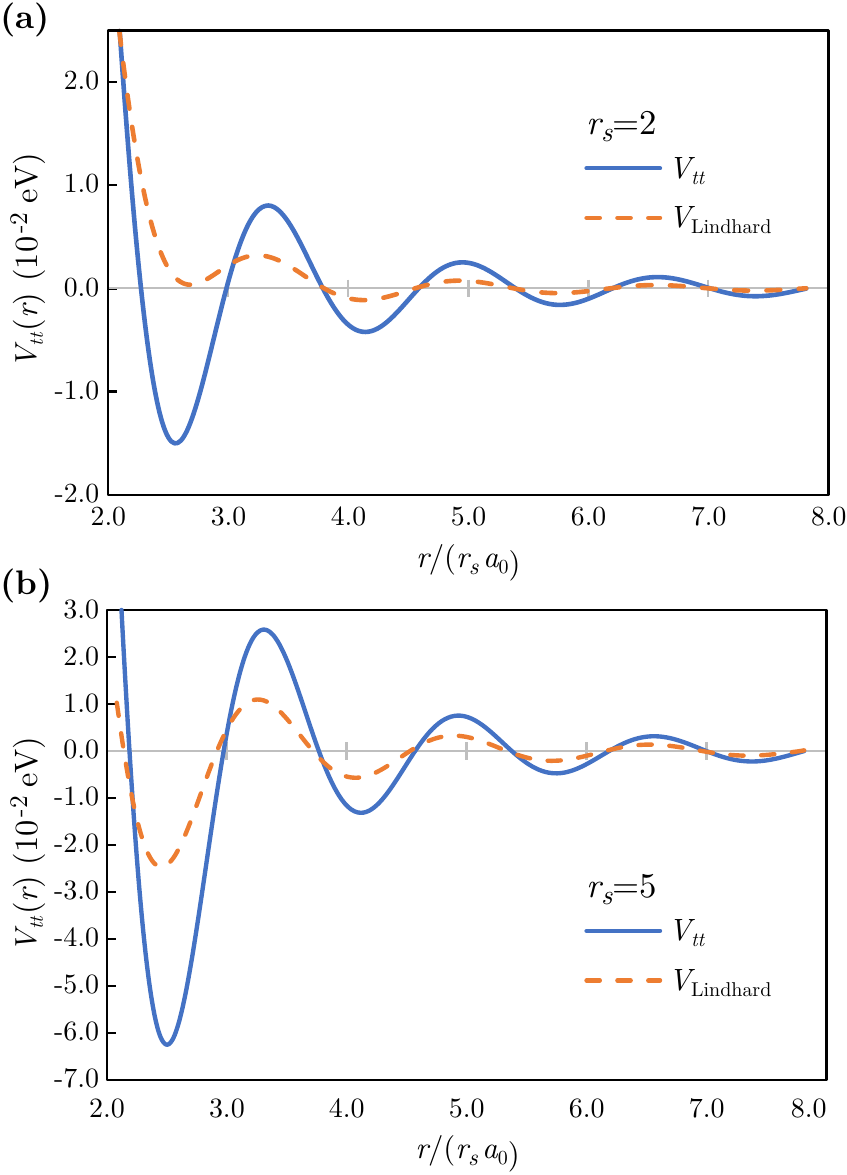}
	\caption{The test charge-test charge interactions for $r_s=2$(a) and 5(b) have been numerically Fourier transformed and are plotted versus distance. The 				Fourier transform of the Lindhard function is shown for comparison. The data are shown from $r/(r_s \, a_0) = 2.2-8$ with a factor of 100 magnification compared to Fig \ref{tc-tc vs r/()|14}}
    \label{tc-tc vs r/()|15}
\end{figure}

Near the compressibility divergence, the vertex correction and thus the dielectric function and $V_{tt}$ are extremely sensitive to small changes, and pressure may be an interesting variable. When applying this formula to real metals, the effective mass and core polarization will re-normalize the equations to make the effective $r_s^\ast$ lower than the actual physical $r_s$. Another Ward identity specifying the renormalization factor $z$ must also be considered.

The Fourier transforms of the test charge-test charge potentials are shown in Fig. \ref{tc-tc vs r/()|14} and Fig. \ref{tc-tc vs r/()|15} compared to the Lindhard potential.

Although it is not shown in Fig. \ref{tc-tc vs r/()|14}, $V_{tt}$ and $V_{\text{Lindhard}}$ converge at small distances $r/(r_s a_0) <0.5$ (derived from $q> 2 \, k_F$) and become equal to the bare interaction at even smaller distances. At intermediate distances (derived from intermediate $q$), $V_{tt}$ is less repulsive than $V_{\text{Lindhard}}$. At large distances the oscillations in $V_{tt}$ are larger than the Friedel oscillations in $V_{\text{Lindhard}}$.  This is dramatically different for $r_s=5$ where there is a broad attractive region around the test charge from $r/(r_s a_0) = 0.6$ to $1.4$. The oscillations at larger distances also have a larger amplitude. 
\begin{figure}[h!]
    \centering
    \includegraphics[width=1.0\columnwidth]{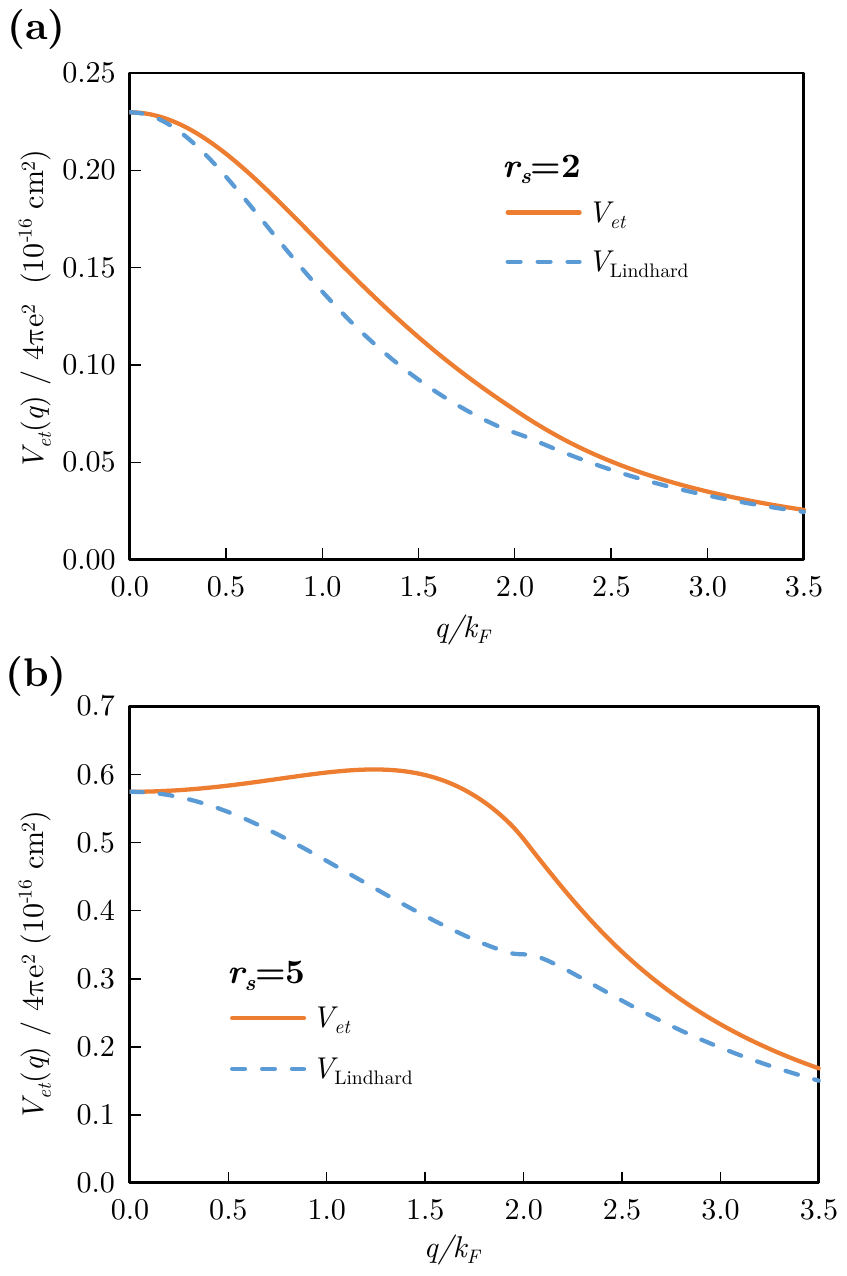}
    \caption{The electron-test charge interaction $V_{et}(q)$ at $r_s=2$ and  $5$. The dashed curve is the Lindhard potential. }
    \label{vet2-vet5}
\end{figure}

The lattice spacing of alkali metals is approximately 1.1 times $r_s$, and the diameter of the core electrons is about 0.5 times $r_s$.  The attractive minimum for $r_s=5$ located at $r/(r_s a_0)=0.8$ has a depth of $0.7 \, eV$ compared to a cohesive energy of rubidium of $0.85 \, eV$.  This attraction can be part of the explanation for the contraction of the inter-atomic spacing in liquid rubidium that is observed in x-ray scattering experiments as a function of pressure and temperature\cite{ref19}. 

The electron-test charge interaction $V_{et}$ is shown in Fig. \ref{vet2-vet5}. The electron-test charge interaction at $q=0$ is equal to $1/q_{TF}^2$ which is the same as the Lindhard and Thomas- Fermi interactions (obtained by setting $G_+=0$). The effect of the vertex correction in the numerator is canceled by the vertex correction in the dielectric function at $q=0$. The effect of exchange and correlation only occurs between zero and about $2.5 \, k_F$. For $r_s=2$, the effects of exchange and correlation are small. The effects are larger for $r_s=5$.
	
\begin{figure}[h!]		
    \begin{center}
    \includegraphics[width=1.0\columnwidth]{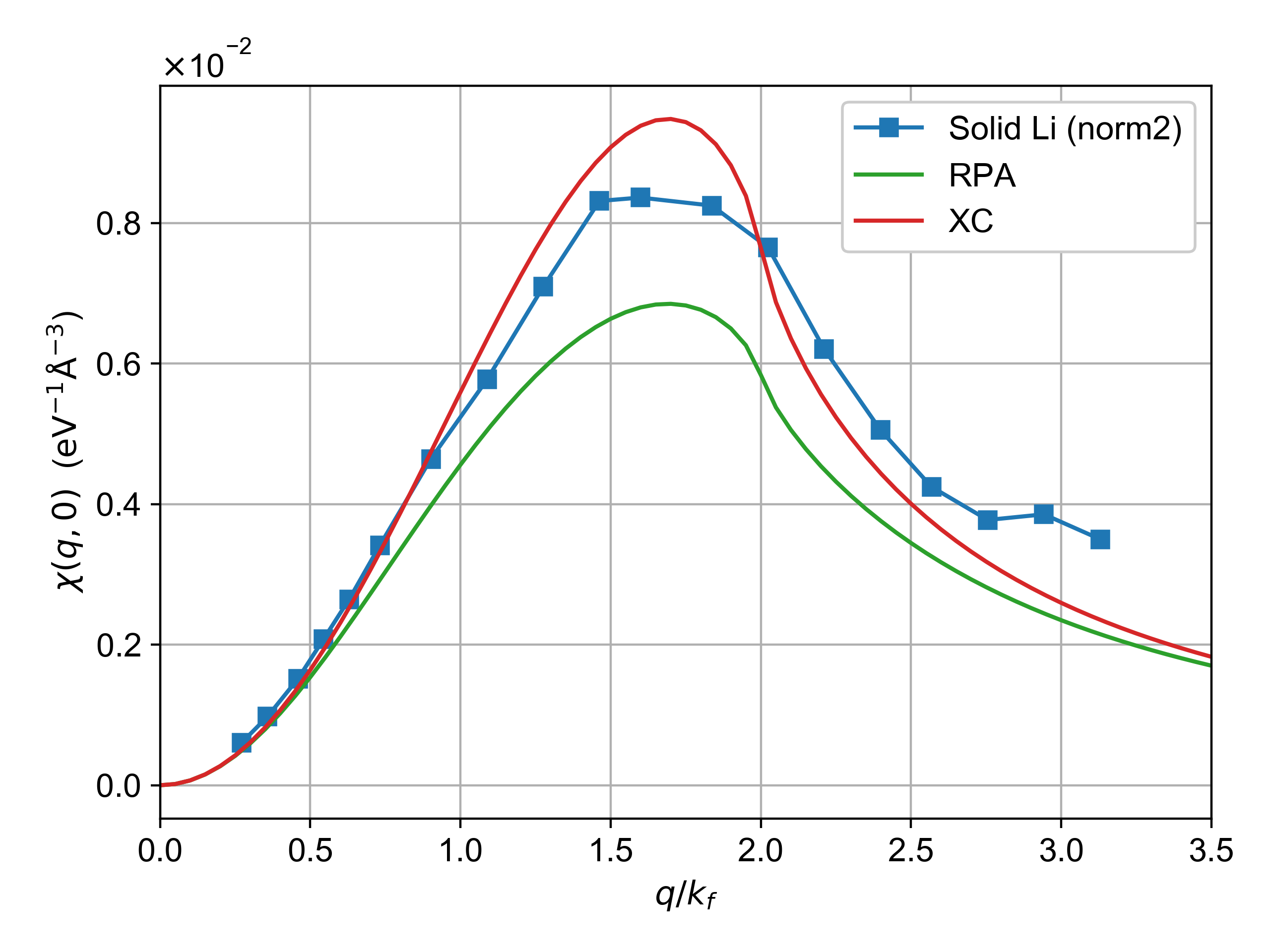}
    \caption{Measured static density response function for lithium from Ref. \cite{ref20} compared to theoretical value with no adjustable parameters. The top curve includes 				exchange and correlation using $G_+(q)$ and the bottom curve uses the Random Phase Approximation which is obtained by setting $G_+(q)=1$. }
    \label{fig16}
    \end{center}
	\end{figure}

A recent paper\cite{ref20} calculated the static density response function of lithium from a Kramers-Kronig transformation of the dynamic structure factor measured by inelastic electron scattering. Figure \ref{fig16} was prepared by T. Hagiya, the lead experimental author, using the formula for $X(q)$ including exchange and correlation provided by us compared to the Random Phase Approximation. 

The density response function is given by:
\begin{eqnarray}
	\chi(q) = \frac{\Pi^0}{(1-G_+ Q)(1+Q/(1-G_+ Q))} = \frac{\lambda\Pi^0}{\varepsilon}	.	\\
	\nonumber	
\end{eqnarray}
The RPA response function is obtained by setting $G_+\!=\!1$.
	
The theory has no adjustable parameters. The experimentalists point out that their data is not accurate enough to definitively distinguish between the response functions using exchange and correlation and the RPA. Nevertheless, the experimental results are very impressive as is the data analysis.

\newpage
\bibliography{reference}
\bibliographystyle{apsrev}
\end{document}